\documentclass[aps,pre,showpacs,noshowkeys,amsmath,amssymb,amsfonts,superscriptaddress,longbibliography,reprint]{revtex4-1}
\usepackage[english]{babel}

\usepackage{graphicx}
\usepackage{bm}
\usepackage{physics}
\usepackage{mathtools}
\usepackage{gensymb}
\allowdisplaybreaks[1]
\usepackage{caption}
\usepackage{subcaption}
\DeclareCaptionLabelSeparator{bar}{~\rule[-0.4ex]{0.2ex}{1em}~}
\DeclareCaptionLabelFormat{subfor}{\textbf{#2}}
\newcommand*\bfcaption[2]{\caption[#1]{\textbf{#1.}#2}}
\usepackage{xcolor}
\definecolor{UBcolor}{HTML}{007CC1}

\usepackage[colorlinks=true,pdfnewwindow=true,linkcolor=UBcolor,citecolor=UBcolor,urlcolor=UBcolor,breaklinks=true,linktocpage]{hyperref}
\usepackage[all]{hypcap}
\usepackage[nameinlink,capitalise]{cleveref}
\makeatletter
\AddToHook{cmd/appendix/before}{\def\cref@section@alias{appendix}\def\cref@subsection@alias{appendix}}
\makeatother
\begin{document}

\title{Active screws: Emergent active chiral nematics of spinning self-propelled rods}

\author{Debarghya Banerjee}
\affiliation{Institute for Theoretical Physics, Georg-August-Universit\"{a}t G\"{o}ttingen, 37077 G\"{o}ttingen, Germany}

\author{Lauritz Hahn}
%\altaffiliation{These authors contributed equally to this work.}
\affiliation{Max Planck Institute for the Physics of Complex Systems, N\"{o}thnitzerst. 38, 01187 Dresden, Germany}
\affiliation{Center for Systems Biology Dresden, Pfotenhauerst. 108, 01307 Dresden, Germany}

\author{Ricard Alert}
%\altaffiliation{These authors contributed equally to this work.}
\email{ricard.alert@ub.edu}
\affiliation{Max Planck Institute for the Physics of Complex Systems, N\"{o}thnitzerst. 38, 01187 Dresden, Germany}
\affiliation{Center for Systems Biology Dresden, Pfotenhauerst. 108, 01307 Dresden, Germany}
\affiliation{Cluster of Excellence Physics of Life, TU Dresden, 01062 Dresden, Germany}
\affiliation{Departament de F\'{i}sica de la Mat\`{e}ria Condensada, Universitat de Barcelona, Barcelona, Spain}
\affiliation{Universitat de Barcelona Institute of Complex Systems (UBICS), Barcelona, Spain}
\affiliation{Instituci\'{o} Catalana de Recerca i Estudis Avan\c{c}ats (ICREA), Barcelona, Spain}

\date{\today}

\begin{abstract}
Several types of active agents self-propel by spinning around their propulsion axis, thus behaving as active screws. Examples include cytoskeletal filaments in gliding assays, magnetically-driven colloidal helices, and microorganisms such as the bacterium \textit{M. xanthus}. Here, we develop a model for spinning self-propelled rods on a substrate, and we coarse-grain it to derive the corresponding hydrodynamic equations. If the rods propel purely along their axis, they form an active nematic at high density and activity. However, spinning rods can also roll sideways as they move. We find that this transverse motion turns the system into a chiral active nematic. Thus, we identify a mechanism whereby individual chirality can give rise to collective local chiral flows. Finally, we analyze experiments on \textit{M. xanthus} colonies to show that they exhibit chiral flows around topological defects, with a chiral activity about an order of magnitude weaker than the achiral one. Our work reveals the collective behavior of active screws, which is relevant to colonies of social bacteria and groups of unicellular parasites.
\end{abstract}

\maketitle

\section{Introduction}

Across scales, Nature has taken advantage of the principle of a screw to drive motion. At the molecular scale, in so-called gliding assays, both microtubules and actin filaments are propelled and spinned by carpets of molecular motors \cite{Meissner2024}. At the cellular scale, rod-shaped bacteria like \textit{Myxococcus xanthus} and \textit{Flavobacterium johnsoniae} glide on surfaces by spinning around their body axis \cite{Wadhwa2022,Mignot2007,Balagam2014,Islam2015,Faure2016,Nakane2013,Shrivastava2015,Shrivastava2015a,Shrivastava2016}. Other bacteria, like \textit{Treponema primitia} swim by spinning their helical bodies \cite{Wadhwa2022}. At larger scales, starfish embryos \cite{Tan2022} and algae like \textit{Chlamidomonas reinhardtii} \cite{Cortese2021} and \textit{Volvox} \cite{Drescher2009} also swim while spinning. Beyond the living world, screw-like propulsion was achieved in synthetic helical micropropellers driven by magnetic fields \cite{Zhang2009,Ghosh2009,Palagi2018} or chemical reactions \cite{Gibbs2015,Palagi2018}, and it was also found in simulations of active droplets \cite{Tjhung2017a,Carenza2019a}.
%Bacteria of the Spirochaetes phylum swim by spinning their helical bodies.
All these systems share a key feature: Self-propulsion and spinning share the same axis, and the self-propulsion speed is proportional to the spinning rate. These systems therefore behave as active screws. 

\begin{figure}[tb]
\begin{center}
\includegraphics[width=0.9\columnwidth]{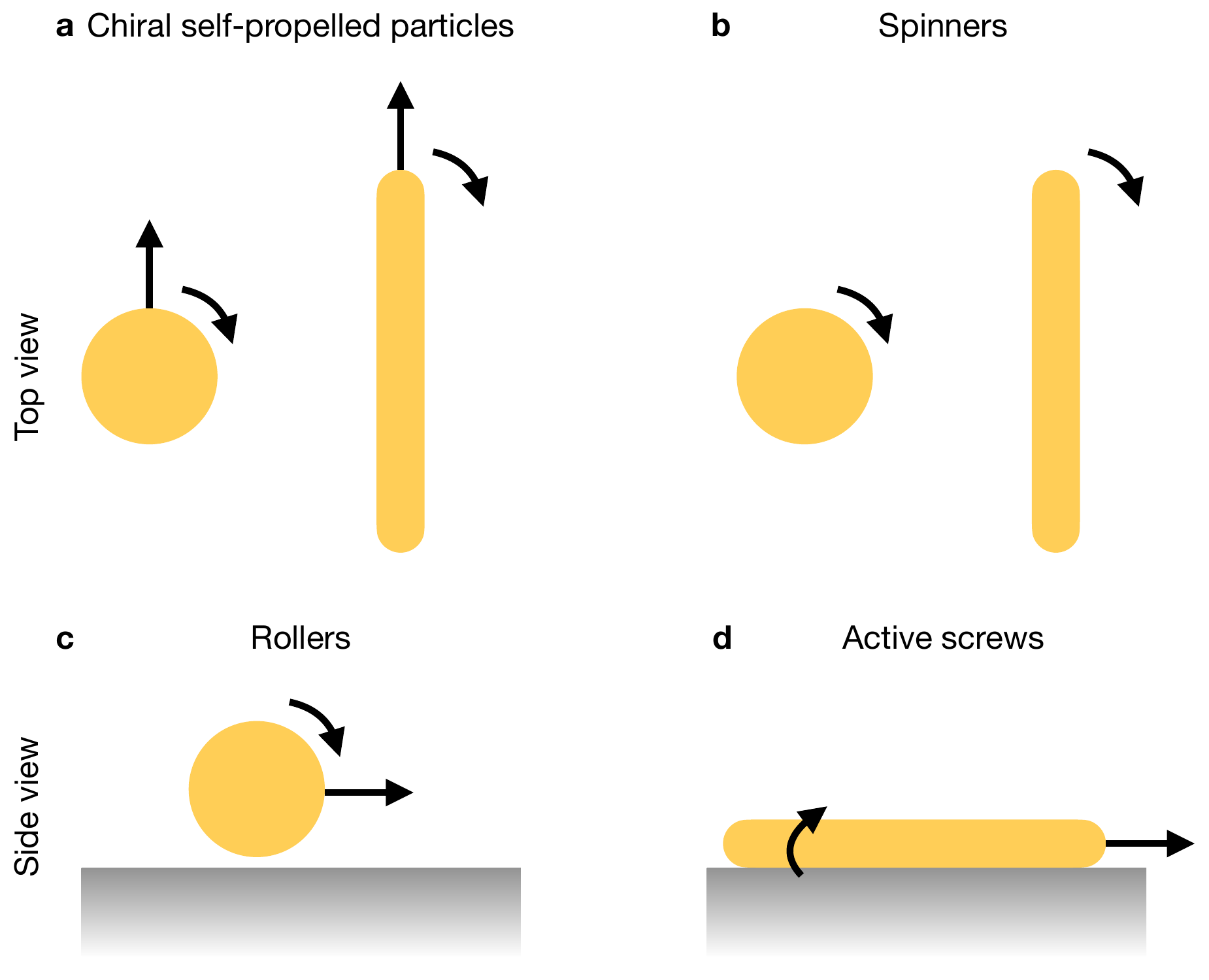}
\end{center}
  {\phantomsubcaption\label{Fig chiral-self-propelled}}
  {\phantomsubcaption\label{Fig spinners}}
  {\phantomsubcaption\label{Fig rollers}}
  {\phantomsubcaption\label{Fig screws}}
\bfcaption{Classes of active chiral particles}{ \subref*{Fig chiral-self-propelled}, Chiral self-propelled particles propelling on the plane and spinning around the axis perpendicular to it. \subref*{Fig spinners}, Spinners do not translate. \subref*{Fig rollers}, Rollers spin around an in-plane axis perpendicular to their self-propulsion. \subref*{Fig screws}, Active screws spin and self-propel along the same axis.} \label{Fig 1}
\end{figure}

In contrast, most work on active chiral particles has focused on objects on a plane that spin around the axis perpendicular to the plane \cite{Bechinger2016,Liebchen2017,Liebchen2022,Bowick2022} (\cref{Fig chiral-self-propelled,Fig spinners}). Systems with this type of chiral motion include bacterial and sperm cells close to surfaces \cite{Riedel2005,Lauga2006,Petroff2015,Li2024d,Aranson2022}, microtubules \cite{Kim2018}, FtsZ filaments \cite{Dunajova2023}, as well as L-shaped self-propelled colloids \cite{Kummel2013} and magnetic spinners \cite{Grzybowski2000,Soni2019a}. Another class of spinning active particles are rollers, such as Quincke rollers \cite{Bricard2013}, which spin around an axis in the same plane but perpendicular to the self-propulsion axis (\cref{Fig rollers}). In all these cases, the axes of rotation and translation are different. Here, we introduce \emph{active screws} as a class of active particles for which the axes of rotation and translation are the same (\cref{Fig screws}).

Rotation and translation along the same axis are common in microswimmers, for example in bacteria propelled by helical flagella. In that case, however, the swimmer is torque-free, and hence the rotation of the flagella is compensated by a counter-rotation of the cell body \cite{Wadhwa2022,Elgeti2015,Lauga2016}. The swimmer therefore experiences no net rotation, and it exerts a torque dipole. In contrast, we focus on active screws moving on surfaces, such as gliding bacteria, which rotate their entire body in the same direction and are therefore driven by a torque monopole. Recent work started to describe the intracellular flows that power this type of gliding motility \cite{Lettermann2024,Hueschen2024} and to analyze single-cell trajectories \cite{Lettermann2025,Lettermann2026,Vilfan2026}. However, the collective behavior of active screws is only beginning to be addressed \cite{Patra2022,Athani2025,Chahal2026}.
%the exception of vortices formed by crescent-shaped malaria parasites \cite{Patra2022} \ra{and global rotations of microtubules in gliding assays \cite{Athani2025}}, the collective behavior of active screws remains largely unaddressed.

Here, we develop a microscopic model of active screws and we coarse-grain it to obtain the continuum equations that capture their collective behavior. Our results show that, if active screws propel purely along their axis, they form an active nematic. The theory of active nematics was recently used to explain cell flows and the formation of cell layers in colonies of \textit{M. xanthus} \cite{Copenhagen2021,Han2025}. However, the connection between the continuum equations and the microscopic dynamics of this gliding bacterium was not established. Our results provide this connection: We give the coefficients of the continuum theory in terms of the microscopic parameters. Finally, we generalize our findings to active screws that also roll sideways, thus adding a transverse component of motion. In this case, we obtain the equations of chiral active nematics. We thus identify a mechanism whereby the chirality of individual constituents can give rise to collective chiral effects. Overall, our work generalizes the theory of self-propelled rods \cite{Marchetti2013,Bar2020} to spinning particles, and it can be tested in colonies of gliding bacteria.

\section{Microscopic model of active screws} \label{model}

\begin{figure}[tb]
\begin{center}
\includegraphics[width=\columnwidth]{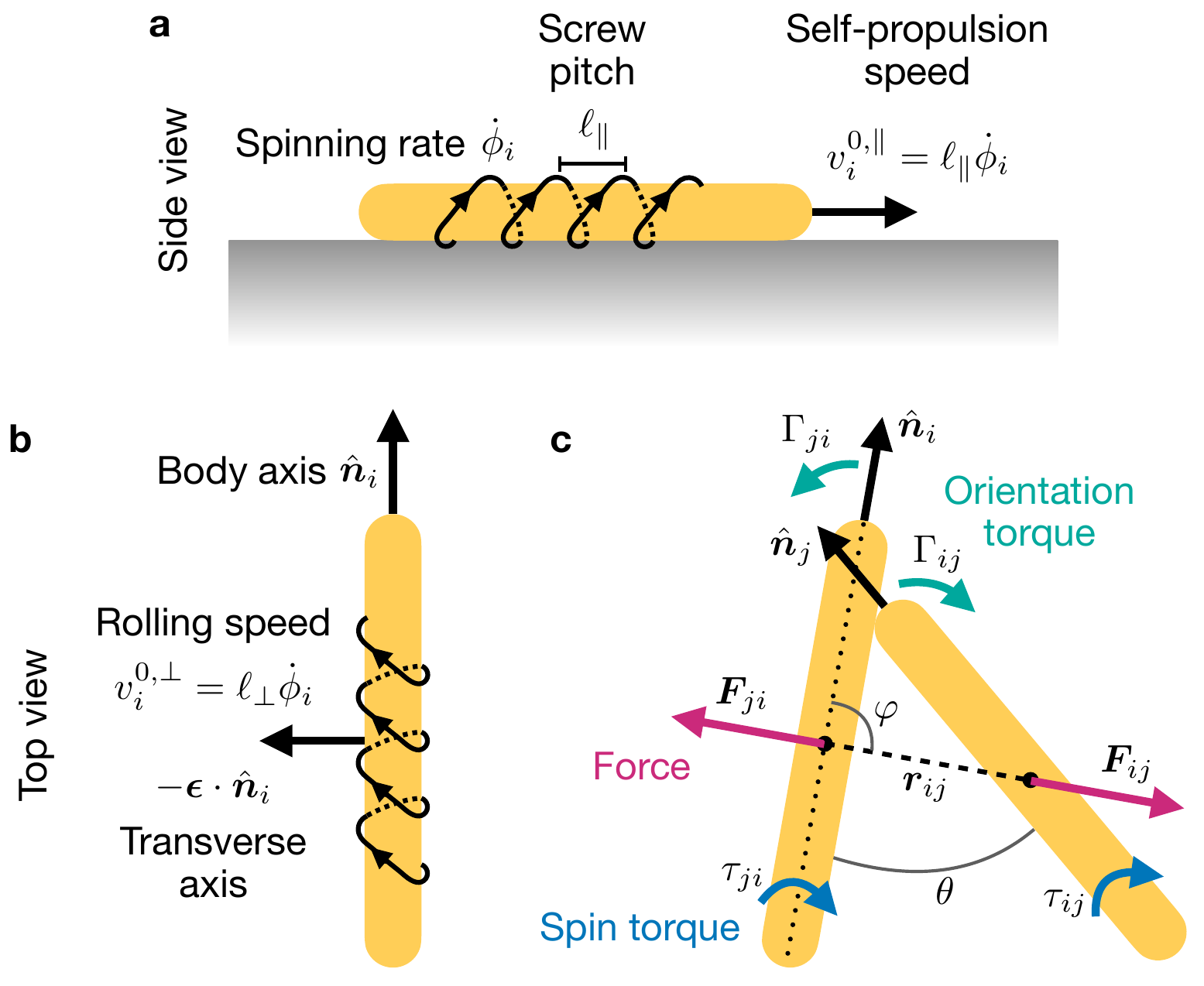}
\end{center}
  {\phantomsubcaption\label{Fig screwing}}
  {\phantomsubcaption\label{Fig rolling}}
  {\phantomsubcaption\label{Fig interactions}}
\bfcaption{Motion and interactions of active screws}{ \subref*{Fig screwing}, An active screw spinning at a rate $\dot\phi_i$ with a pitch $\ell_\parallel$ propels along its axis at a speed $v_i^{0,\parallel} = \ell_\parallel \dot\phi_i$. \subref*{Fig rolling}, An active screw can also propel sideways by rolling at a speed $v_i^{0,\perp}=\ell_\perp \dot\phi_i$. \subref*{Fig interactions}, The interactions between active screws include forces, torques that change their orientations, and torques that change their spinning rate.} \label{Fig 2}
\end{figure}

We propose a model of active screws as point particles with an intrinsic axis corresponding to the orientation of the rod that they represent. We model the spinning of active screws via their spinning rate $\dot\phi_i$ as (\cref{Fig screwing}):
\begin{equation} \label{eq spin}
\dot\phi_i = \omega \Phi_i(t) + \frac{\tau_i}{\xi_\text{s}}.
\end{equation}
The first term accounts for self-spinning at a rate $\omega$. Moreover, gliding bacteria experience stochastic spinning reversals \cite{Herrou2020}. We include them via a dichotomous noise $\Phi_i(t)$ that switches between $+1$ and $-1$ with Poisson statistics, such that $\langle \Phi_i(t) \Phi_j(t') \rangle = \delta_{ij} \exp(-2 f_\text{rev} |t - t'|)$, where $f_\text{rev}$ is the average reversal frequency. Similar to a recent model for eukaryotic cell doublets \cite{Vagne2025}, the second term accounts for spin torque transfer between neighboring screws (blue in \cref{Fig interactions}), $\tau_i = \sum_{j\neq i} \tau_{ji}$, which affects the spinning rate through the spin friction coefficient $\xi_\text{s}$. We assume a torque transfer proportional to the relative velocity between the rod surfaces, and hence to the sum of their spinning rates: $\tau_{ij} = \tau_r(r) (\dot\phi_i + \dot\phi_j) \cos\theta$, where $r=|\bm{r}_{ij}|$ is the distance and $\cos\theta = \hat{\bm{n}}_i \cdot \hat{\bm{n}}_j$ is the projection between the two rods (\cref{Fig interactions}). By affecting the spinning rate, this torque can make neighboring particles either speed up or slow down depending on whether they spin in the opposite or the same direction, respectively.

We then describe screw motion by coupling the spinning rate to translation, such that the self-propulsion speed of rod $i$ along its axis is $v^{0,\parallel}_i = \ell_\parallel \dot{\phi}_i$. Here, $\ell_\parallel$ is the pitch of the screw, i.e., the distance it advances in one revolution (\cref{Fig screwing}). This relation assumes that the screws transduce all the spinning into propulsion, without slipping. This assumption is consistent with measurements on gliding bacteria, where the transduction of spinning into propulsion relies on cell-substrate binding provided by proteins called adhesins \cite{Wadhwa2022,Faure2016,Mignot2007,Islam2015,Nakane2013}. We also allow the rods to roll sideways as a result of their spinning, moving at a speed $v^{0,\perp}_i = \ell_\perp \dot{\phi}_i$ perpendicularly to the rod axis $\hat{\bm{n}}_i \equiv (\cos\theta_i, \sin\theta_i)$, where $\ell_\perp$ is the rolling pitch (\cref{Fig rolling}). If rolling occurs without slipping, then $\ell_\perp = 2\pi R$, with $R$ the rod radius. Thus, the rod positions $\bm{r}_i$ follow
\begin{equation} \label{eq translation}
\frac{\dd \bm{r}_i}{\dd t} = \ell_\parallel \dot{\phi}_i \hat{\bm{n}}_i - \ell_\perp \dot{\phi}_i \bm{\epsilon}\cdot \hat{\bm{n}}_i + \frac{\bm{F}_i}{\xi_\text{t}},
\end{equation}
where $\bm{\epsilon}$ is the Levi-Civita symbol, and $\bm{F}_i = \sum_{j\neq i} \bm{F}_{ji}$ is the force of interaction with neighboring rods, which changes rod positions through the translational friction coefficient $\xi_\text{t}$. We assume the force between two rods to be central, and proportional to their projection: $\bm{F}_{ij} = F_r(r) \cos\theta\, \hat{\bm{r}}_{ij}$ (purple in \cref{Fig interactions}). Brownian motion is typically negligible in gliding cells since they are well attached to the substrate. Therefore, we ignore translational noise.

Finally, the orientation angle of the rods evolves as
\begin{equation} \label{eq orientation}
\frac{\dd \theta_i}{\dd t} = \frac{\Gamma_i}{\xi_\text{r}} + \eta_i(t),
\end{equation}
where $\Gamma_i = \sum_{j\neq i} \Gamma_{ji}$ is the orientation torque from interactions with other rods (green in \cref{Fig interactions}), which affects angle through the rotational friction coefficient $\xi_\text{r}$, and $\eta_i(t)$ is a Gaussian noise with strength given by the rotational diffusivity $D_\text{r}$: $\langle \eta_i(t) \eta_j (t') \rangle = 2 D_\text{r} \delta(t-t') \delta_{ij}$. We assume nematic orientational interactions: $\Gamma_{ij} = \Gamma_r(r) \sin(2\theta)$ (\cref{Fig interactions}). Unlike for other classes of active chiral particles (\cref{Fig chiral-self-propelled,Fig spinners}), here we introduce no self-rotation in the orientation angle $\theta_i$.

\section{Emergence of nematic order} \label{emergence}

\begin{figure}[tbp!]
\begin{center}
\includegraphics[width=0.75\columnwidth]{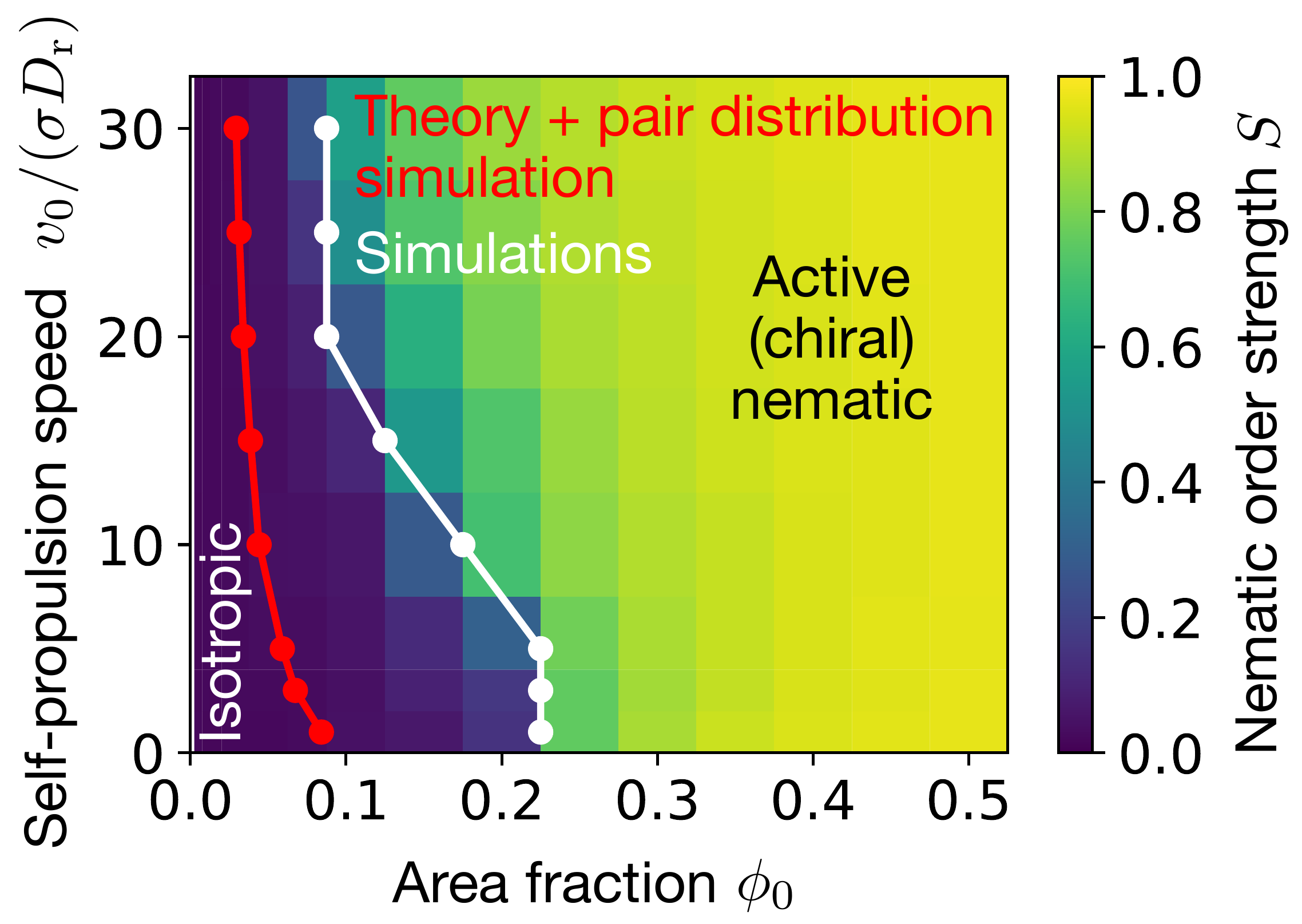}
\end{center}
\bfcaption{Activity-induced nematic order}{ Phase diagram displaying the nematic order strength $S$ (see \cref{simulation-results}). In the simulations, the self-propulsion speed is given by $v_0 = \ell_\parallel \omega$. Following Ref. \cite{Das2024}, the numerical phase boundary (white) is identified from the points of steepest ascent of the measured $S$. The theoretical phase boundary (red) is predicted by setting $a_Q = 0$ in \cref{eq nematic-growth-rate}. The nematic phase is chiral in the presence of rolling.}
\label{Fig simulations}
\end{figure}

To study the emergent phases of active screws, we perform simulations of the Langevin equations of our model (\cref{simulation-details,simulation-results}). For simplicity, we ignore the projection of the repulsive interaction forces. We find that the system transitions from an isotropic to a nematic phase by increasing either the global area fraction of particles $\phi_0$ or the self-propulsion speed $v_0$ (\cref{Fig simulations}). Therefore, an increase in activity can induce the emergence of nematic order.

This result is consistent with previous works on self-propelled rods, which found that activity decreases the threshold for the isotropic-nematic transition \cite{Kraikivski2006,Baskaran2008,Marchetti2013}. This result, however, contrasts with most models of active nematics, for which nematic order is incorporated directly in the free energy as it is assumed to arise from the passive alignment interactions between the constituents \cite{Marchetti2013}. Moreover, our system is dry, meaning that friction completely screens hydrodynamic interactions. Thus, the possibility for activity-induced nematic order in our system is different from a recently-predicted instability whereby isotropic wet active nematics can order due to a feedback between active flows and nematic order \cite{Santhosh2020}.

To predict the isotropic-nematic transition, we coarse-grain the microscopic Langevin equations \crefrange{eq spin}{eq orientation}. First, we eliminate \cref{eq spin} by introducing it into \cref{eq translation}. The stochastic reversals encoded in $\Phi_i(t)$ give rise to an effective rotational diffusivity $D_\text{r}^\text{eff} = D_\text{r} + 2 f_\text{rev}$ on time scales $t\gg f_\text{rev}^{-1}$ \cite{Liu2019}. Second, for this reduced set of Langevin equations, we write the corresponding Smoluchowski equation and we break it into the Bogoliubov-Born-Green-Kirkwood-Yvon (BBGKY) hierarchy to obtain an equation for the one-particle distribution function (\cref{derivation}). From it, we obtain the following equation for the nematic order-parameter tensor,
\begin{equation} \label{eq nematic-order}
\partial_t \bm{Q} =  a_Q [\rho] \, \bm{Q}
\end{equation}
to lowest order in gradients (\cref{isotropic-nematic}). Here,
\begin{equation} \label{eq nematic-growth-rate}
a_Q[\rho] = \frac{\gamma_2}{\pi \xi_\text{r}} \rho - 4D_\text{r}
\end{equation}
is the growth rate of the nematic order. If $a<0$, the isotropic state with $\bm{Q}=\bm{0}$ is stable. If $a>0$, it is unstable to the appearance of nematic order. Thus, the condition $a_Q[\rho] = 0$ determines the phase boundary for the isotropic-nematic transition. The coefficient $\gamma_2$ in \cref{eq nematic-growth-rate} is given by
\begin{equation} \label{eq gamma2}
\gamma_2 = \int_0^{r_\Gamma} \dd r \, r \int_0^{2\pi} \dd\varphi \int_0^{2\pi} \dd\theta \, \Gamma_0 \sin^2(2\theta)  g(r,\varphi,\theta)
\end{equation}
in terms of the torque amplitude $\Gamma_0$ (\cref{simulation-details}) and of the pair correlation function $g(r,\varphi,\theta)$ between two particles separated by a distance vector $\bm{r}$ given by the polar coordinates $(r,\varphi)$ and with a relative orientation $\theta$. We measure this function in simulations (see \cref{simulation-results} and \cref{Fig pairdistribution}) to calculate $\gamma_2$ and thus predict the phase boundary for the isotropic-nematic transition (\cref{Fig simulations}, red), which captures the trend of the numerical results. Note that the isotropic phase, including its correlations, captured by $g(r,\varphi,\theta)$, and its stability, determined by $a_Q[\rho]$, are not affected by rolling, which has consequences only in the nematic phase.

%We then predicted the isotropic-nematic transition using our theory (\cref*{derivation,isotropic-nematic} in \cite{SM}). The resulting phase boundary (light blue in \cref{Fig simulations}) agrees qualitatively with the numerical results.

Moreover, turning off reversals in simulations by setting $f_\text{rev}=0$ slightly shifts the onset of nematic order (\cref{Fig nematic}), and it gives rise to polar flocks (\cref{Fig snapshots}), which produce stronger and longer-lived polarity fluctuations (\cref{Fig polar}), consistent with experiments \cite{Han2025}. At high densities, oppositely-oriented flocks can jam and phase-separate (\cref{Fig snapshots,Fig phase-separation}).

Finally, the nematic phase is locally chiral in the presence of rolling, as we show below. Yet, \cref{eq nematic-order} lacks the term $\sim \Omega \epsilon_{\alpha\gamma}Q_{\gamma\beta}$ describing the global self-rotation of the axis of nematic order introduced in previous works on chiral active nematics \cite{Maitra2019,Maitra2025} and observed in microtubule carpets \cite{Athani2025}. Therefore, systems of active screws do not break chiral symmetry globally, but only locally. This difference is because, in contrast to previous models of chiral active rods (\cref{Fig chiral-self-propelled}), active screws spin around the orientation axis (\cref{Fig screws}), which therefore does not self-rotate.

\section{Active chiral nematic phase} \label{active-chiral}

%To predict the collective behavior of active screws, we coarse-grain the set of Langevin equations \crefrange{eq spin}{eq orientation}. First, we eliminate \cref{eq spin} by introducing it into \cref{eq translation}. However, this does not eliminate the spinning rate $\dot\phi_i$ as a degree of freedom, as the spin torque transfer $\tau_{ij}$ still depends on it. Respectively, the stochastic reversals encoded in $\Phi_i(t)$ give rise to an effective rotational diffusivity $D_\text{r}^\text{eff} = D_\text{r} + 2 f_\text{rev}$ on time scales $t\gg f_\text{rev}^{-1}$ \cite{Liu2019}. Second, for this reduced set of Langevin equations, we write the corresponding Smoluchowski equation and we break it into the Bogoliubov-Born-Green-Kirkwood-Yvon (BBGKY) hierarchy to obtain an equation for the one-particle distribution function $\Psi_1(\bm{r}_1,\theta_1,\dot\phi_1,t)$ (\cref*{derivation} in \cite{SM}). From it, we obtain hydrodynamic equations for the density and polarity fields, $\rho(\bm{r},t)$, $\bm{p}(\bm{r},t)$, respectively (\cref*{derivation} in \cite{SM}), which we present and discuss below. We obtain the equations in the nematic state, in which the angle between rods is $\theta\approx 0$.
Turning to the active chiral nematic phase, we obtain its hydrodynamic equations from our coarse-graining procedure (\cref{nematic-hydrodynamics}). For the rod density field, we obtain
\begin{equation} \label{eq density}
\partial_t \rho = -\bm{\nabla}\cdot\bm{J};\quad \bm{J} = v_\parallel [\rho] \bm{p} - v_\perp \bm{\epsilon}\cdot\bm{p} - \zeta_1 \rho \bm{\nabla}\rho.
\end{equation}
The first term in the flux $\bm{J}$ is advection along the polarity field $\bm{p}(\bm{r},t)$ at a local speed $v_\parallel[\rho] = \omega \ell_\parallel - \zeta_0 \rho$. This speed is produced by the screw motion at a self-spinning rate $\omega_0$, and it decreases with the density field $\rho(\bm{r},t)$. The coefficient $\zeta_0$, given in \cref{eq zeta0}, captures the slowdown resulting from repulsive interactions between rods, which is responsible for motility-induced phase separation \cite{Cates2015,Speck2020}. Torque transfer between neighboring screws, captured by the $\tau_i$ term in \cref{eq spin}, speeds up slower particles and slows down faster ones, such that it has no net effect at the coarse-grained level. The second term in \cref{eq density} corresponds to a flux perpendicular to the polarity at a speed $v_\perp = \omega \ell_\perp$ driven by rolling motion. The transverse speed $v_\perp$ does not have a slowdown effect because we assume that the system is statistically achiral, and hence the repulsive forces from the right and the left of a reference rod cancel out. Finally, the last term in \cref{eq density}, with $\zeta_1$ given in \cref{eq zeta1}, is a diffusive flux induced by repulsive interactions between rods \cite{Zhang2021i}.

The polarity field follows the following equation:
\begin{multline} \label{eq polarity}
\partial_t \bm{p} = - \bm{\nabla}\cdot \left( v_\parallel[\rho] \bm{Q} \right) - \frac{1}{2}\bm{\nabla} \left( v_\parallel[\rho] \rho \right) + \bm{\nabla}\cdot \left( v_\perp \bm{\epsilon}\cdot \bm{Q} \right) \\
+ \frac{1}{2} \bm{\epsilon}\cdot \bm{\nabla} \left( v_\perp \rho \right) + \bm{\nabla}\cdot \left( \zeta_1 \bm{p} \bm{\nabla}\rho \right) - D_\text{r}^\text{eff} \bm{p}.
\end{multline}
The first and second terms are polarity contributions emerging from
%This result shows that, to lowest order in gradients and at times $t\gg 1/D_\text{r}^\text{eff}$, the polarity field has contributions from
the divergence of the nematic order-parameter tensor $\bm{Q}$ and from density gradients. Due to self-propulsion, this polarity then produces density currents in \cref{eq density}, which are a key ingredient of dry active nematics \cite{Ramaswamy2003,Marchetti2013}. In the presence of rolling, with $v_\perp \neq 0$, the polarity gets additional transverse contributions represented by the third and fourth terms in \cref{eq polarity}. Finally, the last term captures the decay of the polarity field due to orientational noise with strength $D_\text{r}^\text{eff}$. This decay is faster than that of the nematic order in \cref{eq nematic-order} due to the contribution of reversal events, which affect the polarity but not the nematic order. Thus, even though both types of order can exist \cite{Grossmann2020}, reversals favor the emergence of nematic in front of polar order.

The density current relates to the velocity field $\bm{v}(\bm{r},t)$ as $\bm{J}=\rho \bm{v}$. For a system with uniform density $\rho=\rho_0$, using the steady-state solution of \cref{eq polarity}, we obtain a velocity field that corresponds to that produced by active stresses of a chiral nematic on a substrate \cite{Furthauer2012,Furthauer2013,Maitra2019,Julicher2018,Maitra2025}:
\begin{equation} \label{eq velocity}
\bm{v} = \frac{1}{\xi} \bm{\nabla} \cdot \bm{\sigma}_\text{act};\qquad \bm{\sigma}_\text{act} = - \zeta \bm{Q} + \zeta_\text{c} \bm{\epsilon}\cdot \bm{Q}.
\end{equation}
Here, following the convention in the field, we take $\bm{Q}$ to be the dimensionless nematic order parameter, whereas $\bm{Q}$ in \cref{eq polarity} is its density, with units of squared inverse length. In \cref{eq velocity}, $\xi$ is the friction coefficient, and $\zeta$ and $\zeta_\text{c}$ are the coefficients for achiral and chiral active stresses, respectively. Our results relate these coefficients to the parameters of the active-screw model as
\begin{equation} \label{eq active-parameters}
\frac{\zeta}{\xi} = \frac{ v_\parallel^2 [\rho_0] - v_\perp^2 }{ D_\text{r}^\text{eff} },\qquad \frac{\zeta_\text{c}}{\xi} = \frac{ 2v_\parallel[\rho_0] \, v_\perp }{ D_\text{r}^\text{eff} },
\end{equation}
with $v_\parallel[\rho_0] = \omega \ell_\parallel - \zeta_0 \rho_0$ as discussed below \cref{eq density}. This result completes our derivation and establishes the connection between the microscopic parameters and hydrodynamic coefficients.

\section{Chiral flows around topological defects}

\begin{figure}[tbp!]
\begin{center}
\includegraphics[width=\columnwidth]{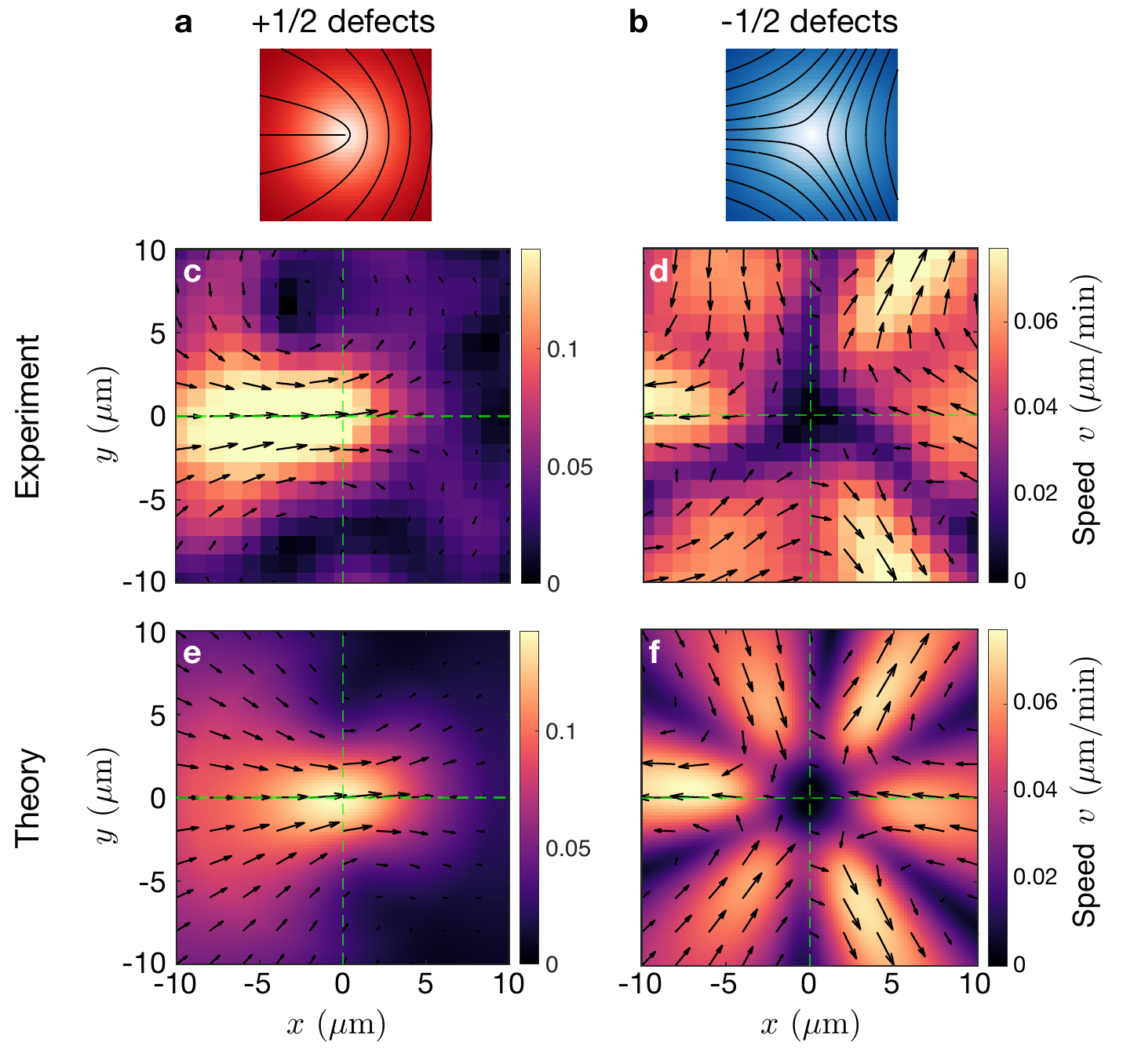}
\end{center}
  {\phantomsubcaption\label{Fig plus-schematic}}
  {\phantomsubcaption\label{Fig minus-schematic}}
  {\phantomsubcaption\label{Fig plus-experiments}}
  {\phantomsubcaption\label{Fig minus-experiments}}
  {\phantomsubcaption\label{Fig plus-theory}}
  {\phantomsubcaption\label{Fig minus-theory}}
\bfcaption{Chiral flows around topological defects}{ \subref*{Fig plus-schematic}-\subref*{Fig minus-schematic}, Defect schematics show the order parameter (color) and the director field (lines). \subref*{Fig plus-experiments}-\subref*{Fig minus-experiments}, Experimental flow fields from Ref. \cite{Han2025}. \subref*{Fig plus-theory}-\subref*{Fig minus-theory}, Fit of active chiral nematic theory (see \cref{fit} and Ref. \cite{Han2025}). Parameter estimates are listed in \cref{t parameters}.}
%The flow fields (bottom) predicted by our theory (\cref{eq velocity}) are slightly chiral. The solutions for the flow field are given in \cite{SM}, and the parameter values are listed in \cref{t parameters}. \ra{Experimental data from Ref. \cite{Han2025}.}}
\label{Fig defects}
\end{figure}

We now compare the predictions of our theory to recent measurements of flow fields around topological defects in colonies of the gliding, spinning bacterium \textit{M. xanthus} \cite{Copenhagen2021,Han2025} (\cref{Fig defects}). Fitting the continuum theory of active nematics to the experimental data gave values of $\zeta/\xi\approx 2$ $\mu$m$^2$/min \cite{Copenhagen2021}. Here, ignoring rolling ($v_\perp=0$), we use independent measurements of the self-propulsion speed \cite{Sabass2017,Liu2019}, $v_\parallel = 1.3$ $\mu$m/min, and of the effective rotational diffusivity \cite{Liu2019}, $D_\text{r}^\text{eff} = 0.34$ min$^{-1}$, to predict $\zeta/\xi \approx 5$ $\mu$m$^2$/min from \cref{eq active-parameters}. This value, predicted from our microscopic model, is slightly larger but comparable with the fit results to experimental data.

\begin{table}[tbp!]
\begin{center}
\begin{tabular}{clc}
Symbol&Description&Estimate\\\hline
$S_0$&Strength of nematic order&$0.992$ \cite{Copenhagen2021}\\
$\ell$&Defect core size&$2.5$ $\mu$m \cite{Copenhagen2021}\\
$\zeta/\xi$&Active stress / friction&$2$ $\mu$m$^2$/min \cite{Copenhagen2021}\\
$\varepsilon$&Friction anisotropy&$0.7$ \cite{Copenhagen2021}\\
$D_\text{r}^\text{eff}$&Effective rotational diffusivity&$0.34$ min$^{-1}$ \cite{Liu2019}\\
$v_\parallel$&Longitudinal self-propulsion speed&$1.3$ $\mu$m/min \cite{Sabass2017,Liu2019}\\
$\zeta_\text{c}/\zeta$&Chiral / achiral active stress&$0.17$\\
%$\zeta_\text{c}/\xi$&Chiral active stress / friction&$0.34$ $\mu$m$^2$/min\\
$v_\perp$&Rolling speed&$0.04$ $\mu$m/min
\end{tabular}
\end{center}
\caption{Parameter estimates for the bacterium \textit{M. xanthus}. The last two estimates are obtained in this work (see text).} \label{t parameters}
\end{table}

Next, we consider the effects of rolling ($v_\perp\neq 0$), which give rise to chiral flows. Chiral flows around defects have been previously proposed to arise in fibrosarcoma cell layers \cite{Hoffmann2020,Yashunsky2022}. Here, we analyzed the experimental data of Ref. \cite{Han2025} and we found that the flows have a non-zero chiral component. The chirality is visible in \cref{Fig plus-experiments,Fig minus-experiments} as the slight upwards flows observed along the $\hat{\bm{x}}$ axis. Fitting the theory of incompressible chiral active nematics to the data (\cref{Fig defects}, see details in \cref{fit}), we found that $|\zeta_\text{c}|/\zeta \approx 0.11$ for $+1/2$ defects and $|\zeta_\text{c}|/\zeta \approx 0.24$ for $-1/2$ defects. From the results of the fits, using \cref{eq active-parameters} and the value of $D_\text{r}^\text{eff}$ listed above, we estimate a rolling speed $v_\perp \approx 0.02-0.06$ $\mu$m/min. All parameter estimates are listed in \cref{t parameters}.

%Here, we analyzed the experimental data in Fig. 4 of Ref. \cite{Copenhagen2021} and we found that the flows have a non-zero chiral component. For defects oriented along the $\hat{\bm{x}}$ axis, as in the schematics in \cref{Fig 3}, chiral flows are revealed by a non-zero average velocity along the $\hat{\bm{y}}$ axis: $\left\langle v_y\right\rangle \neq 0$. For $+1/2$ defects, we compute the average velocity in a circle of radius $R$ centered at the defect. For $R=5$ $\mu$m and $R=10$ $\mu$m, we obtain a ratio of velocity components in the range $\left\langle v_y\right\rangle_R / \left\langle v_x\right\rangle_R \approx 0.15 - 0.19$, showing that the chiral flows are approximately six times weaker than the achiral ones. Solving our model \cref{eq velocity}, even including anisotropic friction \cite{Copenhagen2021,Kawaguchi2017}, gives $\left\langle v_y\right\rangle_R / \left\langle v_x\right\rangle_R = \zeta_\text{c}/\zeta$ \cite{SM}. Hence, we estimate $\zeta_\text{c}/\zeta \approx 0.15 - 0.19$, which, using $\zeta/\xi\approx 2$ $\mu$m$^2$/min \cite{Copenhagen2021,Han2025}, implies $\zeta_\text{c}/\xi \approx 0.30 - 0.38$ $\mu$m$^2$/min. We then use \cref{eq active-parameters} and the values of $v_\parallel$ and $D_\text{r}^\text{eff}$ listed above to obtain a value for the rolling speed $v_\perp \approx 0.04-0.05$ $\mu$m/min. The flow fields predicted by the theory are shown in \cref{Fig 3}, with all the parameter estimates gathered in \cref{t parameters}.

Our analysis revealed chiral flows around topological defects in colonies of \textit{M. xanthus}. While our theory does not attempt to achieve quantitative agreement with the experiments, it identifies a possible mechanism for the chiral flows: The rolling motion of bacteria due to their spinning-based gliding motility. Rolling was achieved in helical microswimmers \cite{Barbot2014}, and is also performed by fruit fly larvae \cite{Liang2025}. Although we know of no direct evidence for rolling in gliding bacteria, our results suggest that chiral cellular flows around defects might be an indirect evidence for this type of cellular motion. Our estimates indicate that the rolling speed could be more than an order of magnitude smaller than the longitudunal self-propulsion speed. Hence, rolling might have been overlooked in the experiments. Based on these findings, future experiments can both search for direct evidence of rolling and investigate alternative mechanisms for chiral flows, including the emergence of polar order during defect unbinding \cite{Beer2025}.

\section{Discussion and outlook}

We introduced a microscopic model for active screws --- a class of active chiral particles that spin around their axis of self-propulsion (\cref{Fig screws}). Compared to other classes of active particles, which are described by their position and orientation, active screws have an additional degree of freedom: their spinning rate. The spinning rate contributes to (i) the speed at which active screws propel themselves forwards, (ii) the speed at which they roll sideways, and (iii) the spin torque transferred among neighboring screws. Whereas previously-studied active chiral particles also spin, they do it perpendicularly to their axis of self-propulsion (\crefrange{Fig chiral-self-propelled}{Fig rollers}). Moreover, in most cases studied so far, the spinning rate was constant, independent of the self-propulsion speed and of interactions with other particles. Hence, the spinning rate was so far a parameter instead of a degree of freedom as in the case of active screws.

We predicted the collective behavior of active screws by coarse-graining the microscopic model. We obtained continuum equations that correspond to a chiral active nematic, albeit with important features such as the absence of active rotation of the nematic axis due to the fact active screws spin around their own axis. Our derivation shows how the microscopic screw and rolling motion give rise to macroscopic achiral and chiral active nematic stresses, respectively.
%These results establish a connection between the parameters of the microscopic model and those of the macroscopic continuum equations.
This micro-macro connection is directly relevant for colonies of gliding bacteria, where single-cell decisions are thought to control collective behaviors \cite{Thutupalli2015,Liu2019,Dinet2021}.
%\ra{For example, our model captures that decreasing the reversal rate leads to longer-lived polarity fluctuations, controlled by the effective rotational diffusivity $D_\text{r}^\text{eff} = D_\text{r} + 2 f_\text{rev}$, consistent with experimental measurements \cite{Han2025}.}

We then analyzed experimental data from Ref. \cite{Copenhagen2021,Han2025} to reveal the presence of chiral flows in colonies of \textit{M. xanthus}. We estimated the corresponding active chiral stresses to be an order of magnitude weaker than the achiral ones. Using the micro-macro connection, we estimated the values of microscopic parameters, such as the rolling speed, from experimental data. Our results could potentially also explain collective chiral flows observed in colonies of other bacteria with screw-like propulsion, such as \textit{C. gingivalis} \cite{Shrivastava2018} and \textit{F. johnsoniae} \cite{Nakane2021}. Overall, using our theory of active screws, we propose that a possible mechanism for emergent chiral flows is the transverse rolling motion that can take place when a screw spins against a substrate. We propose to test these ideas against possible alternative mechanisms in future experiments.

%\cite*{Das2024,Kawaguchi2017}

\bigskip

%\vskip2.5cm

\section*{Acknowledgments}
%\vskip-0.25cm

R.A. thanks Sagnik Garai, Stephan Grill, Pierre Haas, Christina Kurzthaler, David Oriola, and Joshua Shaevitz for discussions. We thank Endao Han and Chenyi Fei for providing experimental data and for assistance with the fits, respectively. R.A. thanks the Isaac Newton Institute for Mathematical Sciences for the support and hospitality during the programme \emph{Anti-diffusive dynamics: from sub-cellular to astrophysical scales}, supported by EPSRC Grant Number EP/R014604/1, when work on this paper was undertaken.

%\vskip2cm

%\bigskip

%\noindent\textbf{Author contributions}
%\vskip0.25cm
%\vskip-0.25cm

%\bigskip

%\noindent\textbf{Competing interests}
%\vskip-0.25cm

%The authors declare no competing interests.

%\bigskip

%\noindent\textbf{Data and code availability}
%\vskip-0.25cm

%Data and code are available \MYhref{https://github.com/dassuchi/Flocking}{on this link}.

\appendix
\renewcommand{\appendixname}{APPENDIX}

\section{SIMULATION DETAILS} \label{simulation-details}

Here, we describe the simulations of our microscopic model. We simulate the dynamics of $N=2500$ active Brownian particles based on \crefrange{eq spin}{eq orientation}.
%the equations
%\begin{align} \label{eq simulation-equations}
%\frac{\dd \bm{r}_i}{\dd t} &= v_0 \hat{\bm{n}}(\theta_i) + \frac{1}{\xi_\text{t}} \sum_{j\neq i} \bm{F}_{ji},\\
%\frac{\dd \theta_i}{\dd t} &= \frac{1}{\xi_\text{r}} \sum_{j\neq i} \Gamma_{ji} + \eta_i(t) + \frac{\pi}{2}\Phi_i(t),
%\end{align}
The particles move in a two-dimensional square box of side $L$ with periodic boundary conditions.
%Here, $\bm{r}_i$ is the position and $\theta_i$ the orientation of particle $i$. Particles propel with speed $v_0$ in the direction $\hat{\bm{n}}_i = (\cos\theta_i,\sin\theta_i)^T$. Particles are also subject to rotational noise $\eta_i(t)$ with diffusivity $D_\text{r}$, such that $\langle \eta_i(t) \eta_j (t') \rangle = 2 D_\text{r} \delta(t-t') \delta_{ij}$. In addition, the particles also undergo stochastic reversals, whereby $\theta_i \rightarrow \theta_i + \pi$, with rate $f_\text{rev}$. We include them through the dichotomous noise $\Phi_i(t)$, which stochastically switches between $\pm 1$ and follows Poisson statistics with $\langle \Phi_i(t) \Phi_j(t') \rangle = \delta_{ij} \exp(-2 f_\text{rev} |t - t'|)$. The prefactor of $\pi/2$ in this noise term ensures that the jumps between $\pm 1$ correspond to jumps of $\pi$ in the orientation angle $\theta_i$.
The particles interact via pairwise forces $\bm{F}_{ji}$, orientational torques $\Gamma_{ji}$, and spin torques $\tau_{ji}$.
%, damped by the translational and rotational friction coefficients $\xi_\text{t}$ and $\xi_\text{r}$, respectively.
To model excluded-volume interactions, we take $\bm{F}_{ji}(\bm{r}_i-\bm{r}_j) = -\bm{\nabla}_{\bm{r}_j} U (\bm{r}_i - \bm{r}_j)$, with $U$ being the Weeks-Chandler-Andersen (WCA) potential,
\begin{equation} \label{eq WCA}
U(\bm{r}) = \left\{ \begin{array}{ll} 4 U_0 \left[ \left(\frac{\sigma}{|\bm{r}|}\right)^{12} - \left(\frac{\sigma}{|\bm{r}|}\right)^6 \right] & \text{if } |\bm{r}|\leq r_\text{WCA},\\
\\
0 & \text{if } |\bm{r}\geq r_\text{WCA},\end{array}\right.
\end{equation}
where $U_0$ is the potential strength, $\sigma$ is the particle diameter, and $r_\text{WCA}$ is the force cutoff distance. For simplicity, in our simulations we ignore the projection $\cos\theta$ of the force between rods that was introduced in \cref{model}. Respectively, to model the nematic alignment between rods, we take a short-range torque $\Gamma_{ji}(\bm{r}_i - \bm{r}_j, \theta_i - \theta_j)$ given by
\begin{equation} \label{eq torque-simulations}
\Gamma_{ji} (\bm{r},\theta) = \left\{ \begin{array}{ll} \Gamma_0 \sin(2\theta) & \text{if } |\bm{r}|\leq r_\Gamma,\\
\\
0 & \text{if } |\bm{r}\geq r_\Gamma,\end{array}\right.
\end{equation}
where $\Gamma_0$ is the orientation torque strength and $r_\Gamma$ is the torque cutoff distance, which we set to $r_\Gamma = 2\sigma$.

For the spin torques in our simulations, we use a modified version of the one presented in \cref{model} to improve numerical stability: Consider two interacting active screws $i$ and $j$ of length $L$ at relative angle $\theta=\theta_j-\theta_i$, spinning at rates $\dot\phi_i$ and $\dot\phi_j$. The overlap of the two particles is given by the projection $L|\cos\theta|$; this is the effective length over which the screws can exert spin torques on each another. Let us now consider the spin torque exerted by screw $j$ on screw $i$, which we assume to be proportional to the effective overlap and the relative surface velocity, which is approximately $\dot \phi_i+ \dot\phi_j\cos(\theta)$. Here, the cosine appears since flipping the orientation of $j$ will flip its spinning axis. As a result, the spin torque reads %Spin torques $\tau_{ji}(\bm{r}_i - \bm{r}_j, \theta_i - \theta_j,\dot\phi_i,\dot\phi_j)$ are computed by projecting the body length and rolling axis of one particle onto those of the other particle:
\begin{equation} \label{eq spintorque-simulations}
\tau_{ji} (\bm{r},\theta,\dot\phi_i,\dot\phi_j) = \left\{ \begin{array}{ll} \tau_0 |\cos(\theta)|(\dot \phi_i+ \dot\phi_j\cos(\theta)) & \text{if } |\bm{r}|\leq r_\tau,\\
\\
0 & \text{if } |\bm{r}\geq r_\tau,\end{array}\right.
\end{equation}
where $\tau_0$ is the spin torque strength and the cutoff $r_\tau=2\sigma$ the same as $r_\Gamma$.

%We do not include rolling in our simulations since a transverse direction can only be defined for rod-shaped particles, not for the point particles used here. Thus, our simulation results will illustrate the transition from isotropic to nematic order in our system. In the presence of rolling, the nematic phase would then have chiral stresses as shown through the coarse-graining results discussed in the Main Text.

We set simulation units by defining the particle diameter $\sigma$ and the inverse rotational diffusivity $D_\text{r}^{-1}$ to be the units of distance and time, respectively. The simulation is performed using an explicit Euler-Mayurama method for the time evolution, with time step $\Delta t = 10^{-5}$, for up to time $t_\text{max} = 150$, and with parameter values $U_0/\xi_\text{t} = 2.5$, $\Gamma/\xi_\text{r} = 10$, $f_\text{rev} = 1$, $\tau_0 = -0.05$, $\xi_\text{s} = 1$, $\ell_\parallel = 6.64\pi$, and $\ell_\perp = 0.23\pi$. To build the phase diagram, we then vary the self-propulsion speed $v_0 = \ell_\parallel \omega$, which we vary by adjusting the spinning rate $\omega$, and the global area fraction $\phi_0 = N \pi (\sigma/2)^2 / L^2$, which we vary by adjusting the system size $L$.

\section{SIMULATION RESULTS} \label{simulation-results}

\begin{figure}[btp!]
\begin{center}
\includegraphics[width=\columnwidth]{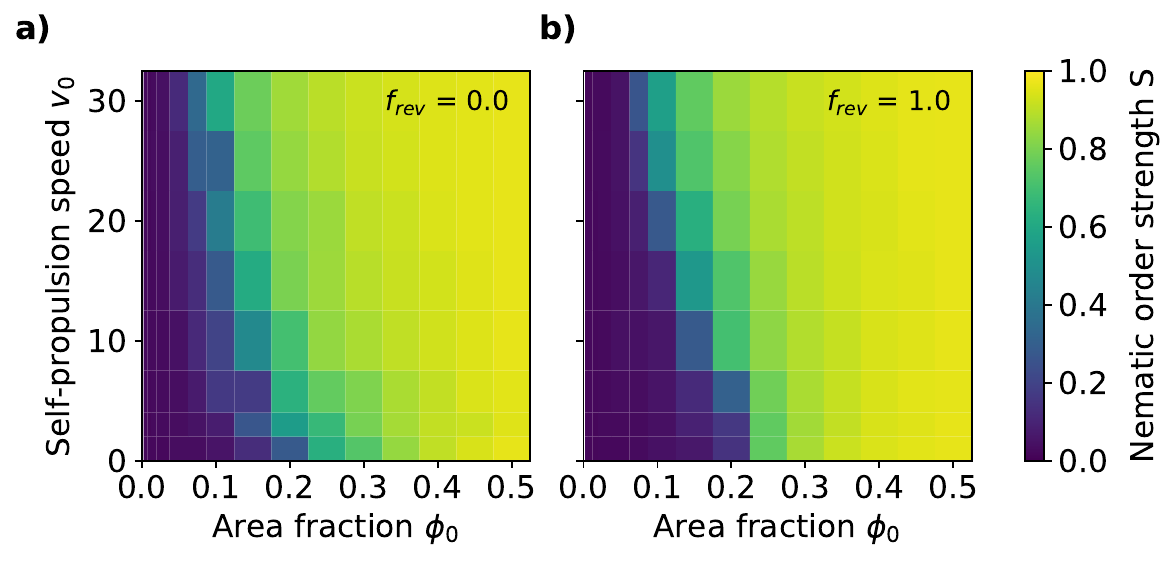}
\end{center}
 {\phantomsubcaption\label{Fig nematic-no-reversals}}
 {\phantomsubcaption\label{Fig nematic-reversals}}
\bfcaption{Nematic order in simulations}{ Phase diagrams showing activity-induced nematic order both without reversals (\subref*{Fig nematic-no-reversals}) and with reversals at a rate $f_\text{rev} =1$ (\subref*{Fig nematic-reversals}).}
\label{Fig nematic}
\end{figure}

After running the simulations to a steady state, we characterize the state of the system as follows. As either the self-propulsion speed $v_0$ and/or the area fraction $\phi_0$ increase, nematic order emerges. We quantify it via the nematic order strength $S = (1/N) \langle | \sum_{j=1}^N e^{i2\theta_j(t)} | \rangle_t$, which we show in \cref{Fig simulations,Fig nematic}.

\begin{figure}[h!]
\begin{center}
\includegraphics[width=\columnwidth]{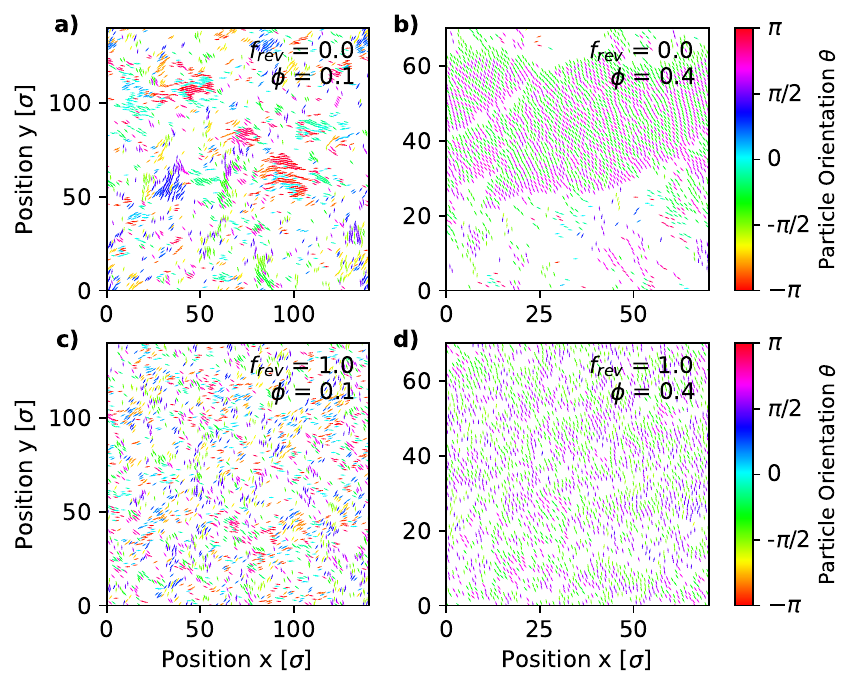}
\end{center}
 {\phantomsubcaption\label{Fig snapshots-f0-phi0.1}}
 {\phantomsubcaption\label{Fig snapshots-f0-phi0.4}}
 {\phantomsubcaption\label{Fig snapshots-f1-phi0.1}}
 {\phantomsubcaption\label{Fig snapshots-f1-phi0.4}}
\bfcaption{Simulation snapshots of active screws with and without polarity reversals}{ Without reversals, active screws form polar clusters at low area fraction (\subref*{Fig snapshots-f0-phi0.1}), whereas at high area fraction, they phase separate into a dilute and a dense nematic phase (\subref*{Fig snapshots-f0-phi0.4}). With reversals, there are no polar clusters at low area fraction (\subref*{Fig snapshots-f1-phi0.1}). Similarly, at high area fraction, there is no phase separation, but the systems achieves a state with nematic order (\subref*{Fig snapshots-f1-phi0.4}). All four snapshots are taken at $v_0=10$ in simulation units (see text).}
\label{Fig snapshots}
\end{figure}

\begin{figure}[h!]
\begin{center}
\includegraphics[width=\columnwidth]{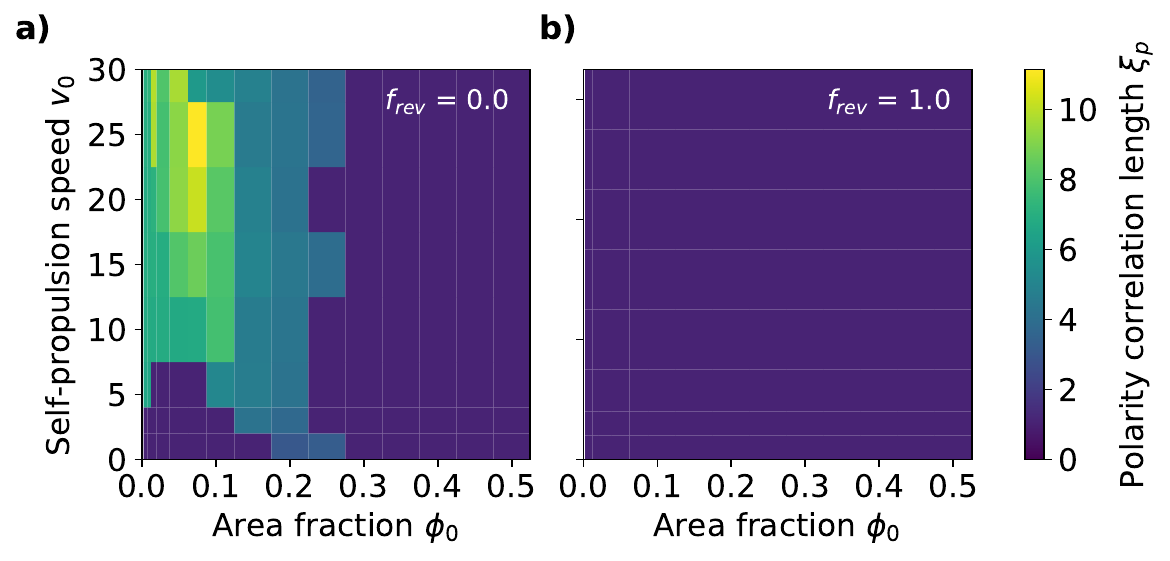}
\end{center}
 {\phantomsubcaption\label{Fig polar-no-reversals}}
 {\phantomsubcaption\label{Fig polar-reversals}}
\bfcaption{Local polar order in simulations}{ Phase diagrams showing the polarity correlation length (see text) both without reversals (\subref*{Fig polar-no-reversals}) and with reversals at a rate $f_\text{rev} =1$ (\subref*{Fig polar-reversals}).}
\label{Fig polar}
\end{figure}

\begin{figure}[tbhp!]
\begin{center}
\includegraphics[width=\columnwidth]{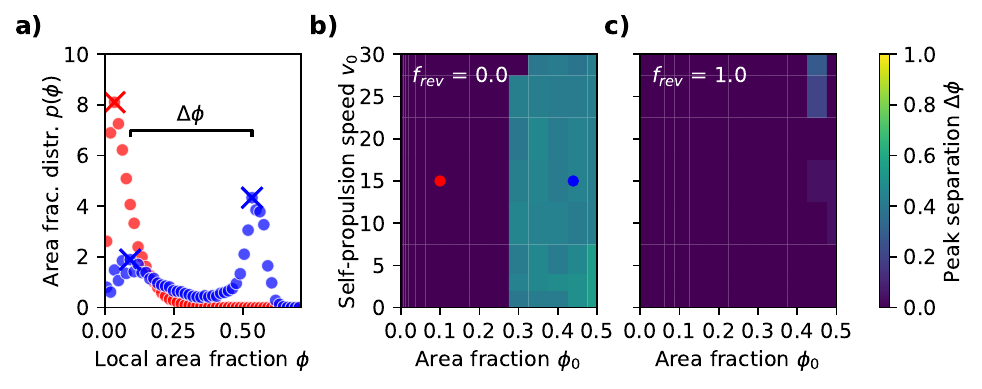}
\end{center}
 {\phantomsubcaption\label{Fig area-fraction-distribution}}
 {\phantomsubcaption\label{Fig phase-separation-no-reversals}}
 {\phantomsubcaption\label{Fig phase-separation-reversals}}
\bfcaption{Phase separation in simulations}{ \subref*{Fig area-fraction-distribution}, Probability distribution of local area fractions. A bimodal distribution, with two peaks (blue), indicates phase separation. \subref*{Fig phase-separation-no-reversals},\subref*{Fig phase-separation-reversals}, Phase diagrams showing the separation in area fraction between the peaks in the area-fraction distribution shown in \subref*{Fig area-fraction-distribution}, both without reversals (\subref*{Fig phase-separation-no-reversals}) and with reversals at a rate $f_\text{rev} =1$ (\subref*{Fig phase-separation-reversals}). If the distribution has a single peak, the peak separation is set to zero.}
\label{Fig phase-separation}
\end{figure}

\begin{figure}[h!]
\begin{center}
\includegraphics[width=\columnwidth]{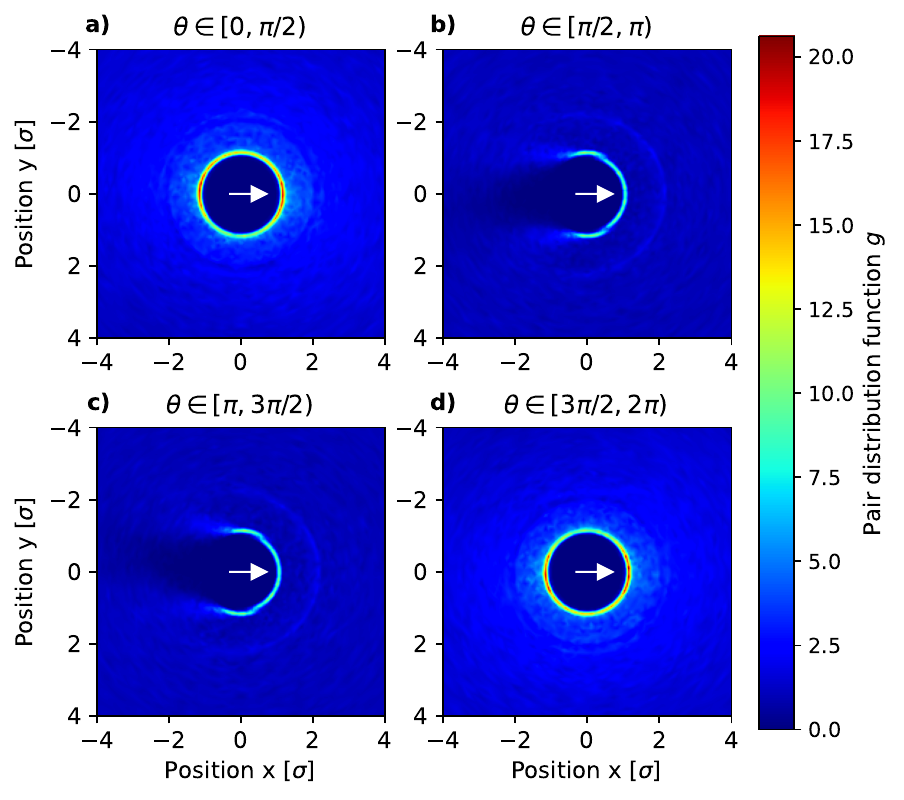}
\end{center}
 {\phantomsubcaption\label{Fig pairdistr-quadrant1}}
 {\phantomsubcaption\label{Fig pairdistr-quadrant2}}
 {\phantomsubcaption\label{Fig pairdistr-quadrant3}}
 {\phantomsubcaption\label{Fig pairdistr-quadrant4}}
\bfcaption{Pair distribution function sampled from simulations}{ Pair distribution function $g(r,\varphi,\theta)$ sampled numerically using \cref{eq g}, which we use to compute the nematic order growth rate in \cref{eq nematic-growth-rate,eq gamma2}. Here, we show $g$ for $\phi=0.05$, $v_0=10$ and $f_\text{rev}=0$, in the isotropic phase. The panels correspond to each of the four quadrants of the relative orientation $\theta$ between two particles, as indicated above the plots. The white arrows indicate the orientation of the reference particle. The pair distribution functions for $f_\text{rev}\neq0$ (not shown here) are qualitatively similar to the ones shown here.}
\label{Fig pairdistribution}
\end{figure}

In the absence of direction reversals, we observe polar clusters (\cref{Fig snapshots-f0-phi0.1}). To characterize the emergence of such local polar order, we obtain the polarity correlation length $\ell_p$ defined by $C_p(r) = \left\langle \bm{p}(\bm{r}) \cdot \bm{p}(\bm{0}) \right\rangle \sim e^{-r/\xi_p}$. We plot $\xi_p$ in \cref{Fig polar}, which shows that reversals readily destroy local polar order by destabilizing polar clusters. Without reversals, at high area fractions, we find that the polar clusters collide often against one another and form aggregates (\cref{Fig snapshots-f0-phi0.4}), which leads to phase separation (\cref{Fig phase-separation}). We quantify phase separation via the distribution of local area fractions, $P(\phi)$ (\cref{Fig area-fraction-distribution}). At low area fraction $\phi_0$, the distribution has a single peak, reflecting density fluctuations of a single phase (\cref{Fig area-fraction-distribution}, red). However, at higher area fractions $\phi_0$, the distribution has two peaks, which correspond to the coexisting dense and dilute phases (\cref{Fig area-fraction-distribution}, blue).

To predict the isotropic-nematic transition using the theory that we develop in \cref{isotropic-nematic} below, we measure the pair distribution function $g(r,\varphi,\theta)$ from our simulations. The pair distribution is a function of the distance vector $\bm{r}$ connecting two particles --- expressed here in polar coordinates as the distance $r$ and the polar angle $\varphi$ --- and of the relative orientation $\theta$ of the two particles. We compute the $g(r,\varphi,\theta)$ as outlined in Ref. \cite{Das2024}:
\begin{equation} \label{eq g}
g(r,\varphi,\theta)=\frac{2\pi \mathcal N(r,\varphi,\theta)}{A(r)Nt_\text{run}\rho\Delta\theta},
\end{equation}
where $\mathcal N(r,\varphi,\theta)$ is a histogram of the number of particles at the given distance, positional and relative orientational angle, which we obtain numerically from our simulations. In the denominator, $A(r)=r\Delta r\Delta\varphi$ is the area of an annular segment of radial width $\Delta r$ and angular opening $\Delta\varphi$ at distance $r$. Here, $\Delta r$, $\Delta\varphi$ and $\Delta \theta$ are the bin sizes for distance, positional and relative orientational angle, respectively. Finally, $N$ is the particle number, $\rho$ the number density of particles, and $t_\text{run}$ the number of snapshots which are used to generate the histogram. To obtain the phase boundaries in \cref{Fig simulations}, we chose bins of size $\Delta r = 0.5$ and $\Delta\varphi = \Delta \theta = \pi/180$, and we averaged over $t_\text{run}=100$ snapshots in simulations with $N=2500$ particles. We show an example of the pair distribution function in \cref{Fig pairdistribution}.

\section{COARSE-GRAINING: FROM THE MICROSCOPIC MODEL TO A HYDRODYNAMIC DESCRIPTION} \label{derivation}

Here, we coarse-grain the microscopic model for active screws, \crefrange{eq spin}{eq orientation}, to derive the hydrodynamic equations describing their collective behavior. We repeat the Langevin equations of motion here for convenience:
%\begin{subequations} \label{eq Langevin}
\begin{align}
\dot\phi_i &= \omega \Phi_i(t) + \frac{\tau_i}{\xi_\text{s}}, \label{eq spin-SI}\\
\frac{\dd \bm{r}_i}{\dd t} &= \ell_\parallel \dot{\phi}_i \hat{\bm{n}}_i - \ell_\perp \dot{\phi}_i \bm{\epsilon}\cdot \hat{\bm{n}}_i + \frac{\bm{F}_i}{\xi_\text{t}}, \label{eq translation-SI}\\
\frac{\dd \theta_i}{\dd t} &= \frac{\Gamma_i}{\xi_\text{r}} + \eta^\text{r}_i(t), \label{eq orientation-SI}
\end{align}
%\end{subequations}
The meaning of the different terms is explained in \cref{model}. To perform the coarse-graining, we follow the derivations in Refs. \cite{Zhang2021i,Das2024} and extend them to include the spinning-rate variable specific to active screws.

\subsection{Eliminating the equation of motion for the spinning rate}

As explained in \cref{emergence}, before starting the coarse-graining, we eliminate \cref{eq spin-SI} by introducing it into \cref{eq translation-SI}. This substitution, however, does not eliminate the spinning rate $\dot\phi_i$ as a degree of freedom, as $\dot\phi_i$ is still present in the spin torque transfer $\tau_i$. In addition to the interaction term represented by $\tau_i$, \cref{eq spin-SI} features the stochastic active driving term $\omega \Phi_i(t)$, which captures stochastic reversals of the self-propulsion direction. Such reversals were shown to give rise to an effective rotational diffusivity $D_\text{r}^\text{eff} = D_\text{r} + 2 f_\text{rev}$, where $f_\text{rev}$ is the average reversal frequency \cite{Liu2019}. Therefore, after eliminating \cref{eq spin-SI}, the effective Langevin equations of motion for active screws reduce to
\begin{align}
\frac{\dd \bm{r}_i}{\dd t} &= \left[\omega + \frac{\tau_i}{\xi_\text{s}} \right] \left[ \ell_\parallel \hat{\bm{n}}_i - \ell_\perp \bm{\epsilon}\cdot \hat{\bm{n}}_i \right] + \frac{\bm{F}_i}{\xi_\text{t}}, \label{eq translation-effective}\\
\frac{\dd \theta_i}{\dd t} &= \frac{\Gamma_i}{\xi_\text{r}} + \eta^\text{r,eff}_i(t), \label{eq orientation-effective}
\end{align}
where $\langle \eta^\text{r,eff}_i(t) \eta^\text{r,eff}_j (t') \rangle = 2 D_\text{r}^\text{eff} \delta(t-t') \delta_{ij}$.

%\bigskip

\subsection{Smoluchowski equation and the BBGKY hierarchy}

The behavior of the system encoded in the set of coupled Langevin equations \cref{eq translation-effective,eq orientation-effective} can be equivalently described by the Smoluchowski equation for the N-particle distribution function $\Psi_N(\bm{r}_1,\hat{\bm{n}}_1,\dot\phi_1,\ldots,\bm{r}_N,\hat{\bm{n}}_N,\dot\phi_N;t)$, which is the probability density of finding the $N$ particles at positions $\bm{r}_1,\ldots,\bm{r}_N$ with orientations $\hat{\bm{n}}_1,\ldots,\hat{\bm{n}}_N$ and spinning at rates $\dot\phi_1,\ldots,\dot\phi_N$ at time $t$:
\begin{equation} \label{eq N-particle}
\partial_t \Psi_N = - \sum_{i=1}^N \left[ \bm{\nabla}_i \cdot \bm{J}_{\text{t},i} + \partial_{\theta_i} J_{\text{r},i}\right].
\end{equation}
Here, $\bm{J}_{\text{t}}$ and $J_{\text{r}}$ are the translational and rotational probability currents, respectively, given by
\begin{subequations}
\begin{align}
\bm{J}_{\text{t},i} &= \left\{ \left[ \omega + \frac{\tau_i}{\xi_\text{s}} \right] \left[ \ell_\parallel \hat{\bm{n}}_i - \ell_\perp \bm\epsilon \cdot \hat{\bm{n}}_i \right] + \frac{\bm{F}_i}{\xi_{\text{t}}}\right\} \Psi_N,\\
J_{\text{r},i} &= \frac{\Gamma_i}{\xi_{\text{r}}}\Psi_N - D_{\text{r}}^\text{eff}\, \partial_{\theta_i}\Psi_N.
\end{align}
\end{subequations}
Note that the two-dimensional particle orientation $\hat{\bm{n}}_i$ is given in terms of the angle $\theta_i$ as $\hat{\bm{n}}_i = (\cos\theta_i,\sin\theta_i)^T$.

Integrating over the positions, orientations, and spinning rates of all particles but one, we obtain an equation for the one-particle distributon function $\Psi_1(\bm{r}_1,\theta_1,\dot\phi_1;t)$:
\begin{multline} \label{eq psi1}
\partial_t \Psi_1 = -\bm{\nabla}_1\cdot\left[ \omega \left( \ell_\parallel \hat{\bm{n}}_1 - \ell_\perp \bm\epsilon \cdot \hat{\bm{n}}_1 \right)  \Psi_1\right] \\
-\bm{\nabla}_1\cdot\left[ \left( \ell_\parallel \hat{\bm{n}}_1 - \ell_\perp \bm\epsilon \cdot \hat{\bm{n}}_1 \right) \frac{\tau_\text{int}}{\xi_\text{s}} \right] - \bm{\nabla}_1 \cdot \frac{\bm{F}_{\text{int}}}{\xi_{\text{t}}} \\
 - \partial_{\theta_1} \frac{\Gamma_{\text{int}}}{\xi_{\text{r}}} + D_{\text{r}}^\text{eff}\, \partial_{\theta_1}^2 \Psi_1.
\end{multline}
Here, $\bm{F}_{\text{int}}$, $\Gamma_{\text{int}}$, and $\tau_\text{int}$ are the collective force, orientational torque, and spin torque, respectively, which encode the effects of interactions on particle 1. For pair-wise interactions, they can be expressed in terms of the two-particle distribution function $\Psi_2 (\bm{r}_1,\theta_1,\dot\phi_1,\bm{r}_2,\theta_2,\dot\phi_2;t)$ as
\begin{widetext}
\begin{subequations} \label{eq collective-psi2}
\begin{align}
\bm{F}_\text{int}(\bm{r}_1,\theta_1,\dot\phi_1;t) &= \int \dd^2\bm{r}' \, \dd\theta' \, \dd\dot\phi' \; \bm{F}(\bm{r}'-\bm{r}_1, \theta'-\theta_1) \,\Psi_2(\bm{r}_1,\theta_1,\dot\phi_1,\bm{r}',\theta',\dot\phi';t),\\
\Gamma_\text{int}(\bm{r}_1,\theta_1,\dot\phi_1;t) &= \int \dd^2\bm{r}'\, \dd\theta' \, \dd\dot\phi' \; \Gamma(\bm{r}'-\bm{r}_1, \theta'-\theta_1)\, \Psi_2(\bm{r}_1,\theta_1,\dot\phi_1,\bm{r}',\theta',\dot\phi';t),\\
\tau_\text{int}(\bm{r}_1,\theta_1,\dot\phi_1;t) &= \int \dd^2\bm{r}'\, \dd\theta' \, \dd\dot\phi' \; \tau(\bm{r}'-\bm{r}_1, \theta'-\theta_1,\dot\phi' + \dot\phi_1)\, \Psi_2(\bm{r}_1,\theta_1,\dot\phi_1,\bm{r}',\theta',\dot\phi';t).
\end{align}
\end{subequations}
\end{widetext}
Here, $\bm{F}(\bm{r},\theta)$, $\Gamma(\bm{r},\theta)$, and $\tau(\bm{r},\theta,\dot\phi)$ are the force, orientation torque, and spin torque of interaction between the two particles. These quantities depend on the relative coordinates $\bm{r} = \bm{r}'-\bm{r}_1$, $\theta = \theta' - \theta$, and $\dot\phi = \dot\phi' + \dot\phi_1$ of the interacting pair. With the assumptions discussed in \cref{model}, the dependencies of the interaction forces and torques on the relative coordinates reduce to
\begin{subequations} \label{eq force-torque-fields-expressions}
\begin{align}
\bm{F}(\bm{r},\theta) &= - F_r(r) \cos\theta \,\hat{\bm{r}}, \label{eq force-field}\\
\Gamma(\bm{r},\theta) &= \Gamma_r(r) \sin(2\theta), \label{eq orientation-torque-field}\\
\tau(\bm{r},\theta,\dot\phi) &= \tau_r(r) \dot\phi \cos\theta, \label{eq spin-torque-field}
\end{align}
\end{subequations}
where the functions $F_r(r)$, $\Gamma_r(r)$, and $\tau_r(r)$ capture unspecified radial dependencies of the interactions, with $r=|\bm{r}|$ being the distance between the interacting particles.

With the collective force and torques given by \cref{eq collective-psi2}, \cref{eq psi1} is an integro-differential equation for $\Psi_1$ that involves $\Psi_2$. Therefore, \cref{eq psi1} is the first equation in the BBGKY hierarchy. To truncate the hierarchy, we decompose $\Psi_2$ as
\begin{multline}
\Psi_2(\bm{r}_1,\theta_1,\dot\phi_1,\bm{r}',\theta',\dot\phi';t) = \Psi_1(\bm{r}',\theta',\dot\phi';t) \\
\times g(\bm{r}',\theta',\dot\phi' | \,\bm{r}_1,\theta_1,\dot\phi_1;t) \Psi_1(\bm{r}_1,\theta_1,\dot\phi_1;t).
\end{multline}
Here, $\Psi_2(\bm{r}_1,\theta_1,\dot\phi_1,\bm{r}',\theta',\dot\phi';t)$ is the density of particle pairs with one particle with position $\bm{r}_1$, orientation $\theta_1$, and spinning rate $\dot\phi_1$, and another particle with position $\bm{r}'$, orientation $\theta'$, and spinning rate $\dot\phi'$ at time $t$. Respectively, $g$ is the dimensionless pair distribution function that encodes the conditional probability of finding a particle with position $\bm{r}'$, orientation $\theta'$, and spinning rate $\dot\phi'$ given that another particle has position $\bm{r}_1$, orientation $\theta_1$, and spinning rate $\dot\phi_1$. Introducing this decomposition into \cref{eq psi1} allows us to express it as a closed equation for $\Psi_1$, hence closing the hierarchy. This closure goes beyond the molecular chaos approximation, as it keeps information about pair correlations in the pair distribution function $g$.

In homogeneous steady states, the probability distributions are time-independent, and the pair correlations do not depend on the coordinates of a given particle but only on the relative coordinates of particle pairs, defined just above \cref{eq force-torque-fields-expressions}. Hence, we express $g$ in terms of the distance $r$ between particles, the angle $\varphi$ formed between the interparticle distance vector $\bm{r}$ and the orientation vector $\hat{\bm{n}}_1$ of particle 1 (defined by $\hat{\bm{n}}_1\cdot \bm{r} =r \cos\varphi$), the relative orientation $\theta$, and the relative spinning rate $\dot\phi$. Therefore, in homogeneous steady states like the isotropic and nematic states of active screws that we analyze, we have
\begin{equation}
g(\bm{r}', \theta', \dot\phi' | \,\bm{r}_1,\theta_1,\dot\phi_1;t) = g(r,\varphi, \theta, \dot\phi).
\end{equation}
With this decomposition, and changing the integration variables to the relative coordinates, the collective force and torques (\cref{eq collective-psi2}) are expressed as
\begin{widetext}
\begin{subequations} \label{eq collective-g}
\begin{align}
\label{eq collective-force-g}
\bm{F}_{\text{int}}(\bm{r}_1,\theta_1,\dot\phi_1) &=  \Psi_1(\bm{r}_1,\theta_1,\dot\phi_1) \int \dd^2\bm{r} \,\dd\theta \, \dd\dot\phi \; \bm{F}(\bm{r},\theta) \Psi_1(\bm{r} + \bm{r}_1,\theta + \theta_1,\dot\phi - \dot\phi_1) \, g(r,\varphi,\theta,\dot\phi),\\
\label{eq collective-orientation-torque-g}
\Gamma_{\text{int}}(\bm{r}_1,\theta_1,\dot\phi_1) &=  \Psi_1(\bm{r}_1,\theta_1,\dot\phi_1) \int \dd^2\bm{r} \,\dd\theta \,\dd\dot\phi\; \Gamma(\bm{r},\theta) \Psi_1(\bm{r} + \bm{r}_1,\theta + \theta_1,\dot\phi - \dot\phi_1)\, g(r,\varphi,\theta,\dot\phi),\\
\label{eq collective-spin-torque-g}
\tau_{\text{int}}(\bm{r}_1,\theta_1,\dot\phi_1) &=  \Psi_1(\bm{r}_1,\theta_1,\dot\phi_1) \int \dd^2\bm{r} \,\dd\theta \,\dd\dot\phi\; \tau(\bm{r},\theta,\dot\phi) \Psi_1(\bm{r} + \bm{r}_1,\theta + \theta_1,\dot\phi - \dot\phi_1)\, g(r,\varphi,\theta,\dot\phi).
\end{align}
\end{subequations}
\end{widetext}

\subsection{Fourier and gradient expansions}

The one-particle distribution function $\Psi_1$ in the integrand of \cref{eq collective-g} introduces non-local dependencies in $\bm{r}_1$, $\theta_1$, and $\dot\phi_1$. To derive local hydrodynamic equations, we perform a Fourier expansion on the orientation angle,
\begin{equation} \label{eq Fourier-expansion}
\Psi_1 (\bm{r}',\theta',\dot\phi') = \frac{1}{2\pi} \sum_{k=0}^\infty e^{-ik\theta'} \,\tilde{\Psi}_{1,k}(\bm{r}',\dot\phi'),
\end{equation}
and then a gradient expansion of the Fourier components on both the position and the spinning rate,
\begin{multline} \label{eq gradient-expansion}
\tilde\Psi_{1,k}(\bm{r}',\dot\phi')\approx \tilde\Psi_{1,k}(\bm{r}_1,\dot\phi_1) + (\bm{r}'-\bm{r}_1) \cdot \bm{\nabla}_{\bm{r}'}\tilde\Psi_{1,k}(\bm{r}_1,\dot\phi_1) \\
+ (\dot\phi' - \dot\phi_1) \, \partial_{\dot\phi'} \tilde\Psi_{1,k}(\bm{r}_1,\dot\phi_1).
\end{multline}
The arguments in these expressions relate to those of $\Psi_1$ in the integrand of \cref{eq collective-g} via the transformations to the relative coordinates:  $\bm{r} = \bm{r}'-\bm{r}_1$, $\theta = \theta' - \theta$, and $\dot\phi = \dot\phi' + \dot\phi_1$.
%Thus, to lowest order in the gradient expansion, we have
%\begin{equation} \label{eq gradient-Fourier-expansion}
%\Psi_1 (\bm{r}',\theta') \approx \frac{1}{2\pi} \sum_{k=0}^\infty e^{-ik\theta'} \,\tilde{\Psi}_{1,k}(\bm{r}_1).
%\end{equation}

\subsubsection{Zeroth-order contributions}

Based on this gradient expansion, we now obtain first the zeroth-order and then the first-order contributions to the collective interaction force and torques in \cref{eq collective-g}. The interaction force is a vector, which we take to be along the interparticle distance axis $\hat{\bm{r}}$, as in \cref{eq force-field}. We then decompose it into components parallel and perpendicular to the orientation of particle 1: $\hat{\bm{r}} = \cos\varphi \,\hat{\bm{n}}_1 + \sin\varphi \, \hat{\bm{t}}_1$, where $\hat{\bm{t}}_1$ is unit vector transverse to $\hat{\bm{n}}_1$, which thus fulfills $\hat{\bm{t}}_1 \cdot \hat{\bm{n}}_1 = 0$. Using this decomposition in \cref{eq collective-force-g}, the zeroth-order contributions of \cref{eq collective-g} are given by
\begin{widetext}
\begin{subequations} \label{eq collective-force-torques-order0}
\begin{align}
F_{\text{int},\parallel}^{(0)} (\bm{r}_1,\theta_1,\dot\phi_1) &=  - \Psi_1(\bm{r}_1,\theta_1,\dot\phi_1) \frac{1}{2\pi} \sum_{k=0}^\infty e^{-ik \theta_1} \, \tilde\Psi_{1,k} (\bm{r}_1,\dot\phi_1) \zeta^\parallel_{0,k}, \label{eq parallel-collective-force}\\
F_{\text{int},\perp}^{(0)} (\bm{r}_1,\theta_1,\dot\phi_1) &=  - \Psi_1(\bm{r}_1,\theta_1,\dot\phi_1) \frac{1}{2\pi} \sum_{k=0}^\infty e^{-ik \theta_1} \, \tilde\Psi_{1,k} (\bm{r}_1,\dot\phi_1) \zeta^\perp_{0,k},\\
\Gamma_\text{int}^{(0)} (\bm{r}_1,\theta_1,\dot\phi_1) &=  \Psi_1(\bm{r}_1,\theta_1,\dot\phi_1) \frac{1}{2\pi} \sum_{k=0}^\infty e^{-ik \theta_1} \, \tilde\Psi_{1,k} (\bm{r}_1,\dot\phi_1) \gamma_{0,k},\\
\tau_\text{int}^{(0)} (\bm{r}_1,\theta_1,\dot\phi_1) &=  \Psi_1(\bm{r}_1,\theta_1,\dot\phi_1) \frac{1}{2\pi} \sum_{k=0}^\infty e^{-ik \theta_1} \, \tilde\Psi_{1,k} (\bm{r}_1,\dot\phi_1) \tau_{0,k},
\end{align}
\end{subequations}
where we have defined the coefficients
\begin{subequations} \label{eq zeroth-order-coefficients}
\begin{align}
\zeta_{0,k}^\parallel &\equiv \int_0^\infty \dd r \, r \int_0^{2\pi} \dd\varphi \int_0^{2\pi} \dd\theta \int_{-\infty}^\infty \dd\dot\phi \, F(r,\varphi,\theta) g(r,\varphi,\theta,\dot\phi) \cos\varphi \, e^{-ik\theta},\\
\zeta_{0,k}^\perp &\equiv \int_0^\infty \dd r \, r \int_0^{2\pi} \dd\varphi \int_0^{2\pi} \dd\theta \int_{-\infty}^\infty \dd\dot\phi \, F(r,\varphi,\theta) g(r,\varphi,\theta,\dot\phi) \sin\varphi \, e^{-ik\theta},\\
\gamma_{0,k} &\equiv \int_0^\infty \dd r \, r \int_0^{2\pi} \dd\varphi \int_0^{2\pi} \dd\theta \int_{-\infty}^\infty \dd\dot\phi \, \Gamma(r,\varphi,\theta) g(r,\varphi,\theta,\dot\phi) e^{-ik\theta},\\
\tau_{0,k} &\equiv \int_0^\infty \dd r \, r \int_0^{2\pi} \dd\varphi \int_0^{2\pi} \dd\theta \int_{-\infty}^\infty \dd\dot\phi \, \tau(r,\varphi,\theta,\dot\phi) g(r,\varphi,\theta,\dot\phi) e^{-ik\theta}.
\end{align}
\end{subequations}
\end{widetext}

\subsubsection{First-order contributions}

We now obtain the first-order contributions to the collective interaction force and torques in \cref{eq collective-g}. We proceed as for the zeroth-order terms, except that we now also split the spatial gradients in components parallel and perpendicular to $\hat{\bm{n}}_1$. Thus, we obtain
\begin{widetext}
\begin{subequations} \label{eq collective-force-torques-order1}
\begin{align}
F_{\text{int},\parallel}^{(1)} (\bm{r}_1,\theta_1,\dot\phi_1) &=  - \Psi_1(\bm{r}_1,\theta_1,\dot\phi_1) \frac{1}{2\pi} \sum_{k=0}^\infty e^{-ik \theta_1} \, \left[ \zeta_{1,k,\parallel}^\parallel \bm{\nabla}_\parallel + \zeta_{1,k,\perp}^\parallel \bm{\nabla}_\perp + \zeta_{1,k,\dot\phi}^\parallel \partial_{\dot\phi} \right] \tilde\Psi_{1,k} (\bm{r}_1,\dot\phi_1), \label{eq parallel-collective-force-order-1}\\
F_{\text{int},\perp}^{(1)} (\bm{r}_1,\theta_1,\dot\phi_1) &=  - \Psi_1(\bm{r}_1,\theta_1,\dot\phi_1) \frac{1}{2\pi} \sum_{k=0}^\infty e^{-ik \theta_1} \, \left[ \zeta_{1,k,\parallel}^\perp \bm{\nabla}_\parallel + \zeta_{1,k,\perp}^\perp \bm{\nabla}_\perp + \zeta_{1,k,\dot\phi}^\perp \partial_{\dot\phi} \right] \tilde\Psi_{1,k} (\bm{r}_1,\dot\phi_1),\\
\Gamma_\text{int}^{(0)} (\bm{r}_1,\theta_1,\dot\phi_1) &=  \Psi_1(\bm{r}_1,\theta_1,\dot\phi_1) \frac{1}{2\pi} \sum_{k=0}^\infty e^{-ik \theta_1} \, \left[ \gamma_{1,k,\parallel} \bm{\nabla}_\parallel + \gamma_{1,k,\perp} \bm{\nabla}_\perp + \gamma_{1,k,\dot\phi} \partial_{\dot\phi} \right] \tilde\Psi_{1,k} (\bm{r}_1,\dot\phi_1),\\
\tau_\text{int}^{(0)} (\bm{r}_1,\theta_1,\dot\phi_1) &=  \Psi_1(\bm{r}_1,\theta_1,\dot\phi_1) \frac{1}{2\pi} \sum_{k=0}^\infty e^{-ik \theta_1} \, \left[ \tau_{1,k,\parallel} \bm{\nabla}_\parallel + \tau_{1,k,\perp} \bm{\nabla}_\perp + \tau_{1,k,\dot\phi} \partial_{\dot\phi} \right] \tilde\Psi_{1,k} (\bm{r}_1,\dot\phi_1),
\end{align}
\end{subequations}
where we have defined the coefficients
\begin{subequations} \label{eq first-order-coefficients}
\begin{align}
\zeta_{1,k,\parallel}^\parallel &\equiv \int_0^\infty \dd r \, r \int_0^{2\pi} \dd\varphi \int_0^{2\pi} \dd\theta \int_{-\infty}^\infty \dd\dot\phi \, F(r,\varphi,\theta) g(r,\varphi,\theta,\dot\phi) \, r \cos^2\varphi \, e^{-ik\theta},\\
\zeta_{1,k,\perp}^\parallel &\equiv \int_0^\infty \dd r \, r \int_0^{2\pi} \dd\varphi \int_0^{2\pi} \dd\theta \int_{-\infty}^\infty \dd\dot\phi \, F(r,\varphi,\theta) g(r,\varphi,\theta,\dot\phi) \, r \cos\varphi \sin\varphi \, e^{-ik\theta},\\
\zeta_{1,k,\dot\phi}^\parallel &\equiv \int_0^\infty \dd r \, r \int_0^{2\pi} \dd\varphi \int_0^{2\pi} \dd\theta \int_{-\infty}^\infty \dd\dot\phi \, F(r,\varphi,\theta) g(r,\varphi,\theta,\dot\phi) \, \dot\phi \cos\varphi \, e^{-ik\theta},\\
\zeta_{1,k,\parallel}^\perp &\equiv \int_0^\infty \dd r \, r \int_0^{2\pi} \dd\varphi \int_0^{2\pi} \dd\theta \int_{-\infty}^\infty \dd\dot\phi \, F(r,\varphi,\theta) g(r,\varphi,\theta,\dot\phi) \, r \sin\varphi \cos\varphi \, e^{-ik\theta},\\
\zeta_{1,k,\perp}^\perp &\equiv \int_0^\infty \dd r \, r \int_0^{2\pi} \dd\varphi \int_0^{2\pi} \dd\theta \int_{-\infty}^\infty \dd\dot\phi \, F(r,\varphi,\theta) g(r,\varphi,\theta,\dot\phi) \, r \sin^2\varphi \, e^{-ik\theta},\\
\zeta_{1,k,\dot\phi}^\perp &\equiv \int_0^\infty \dd r \, r \int_0^{2\pi} \dd\varphi \int_0^{2\pi} \dd\theta \int_{-\infty}^\infty \dd\dot\phi \, F(r,\varphi,\theta) g(r,\varphi,\theta,\dot\phi) \, \dot\phi \sin\varphi \, e^{-ik\theta},\\
\gamma_{1,k,\parallel} &\equiv \int_0^\infty \dd r \, r \int_0^{2\pi} \dd\varphi \int_0^{2\pi} \dd\theta \int_{-\infty}^\infty \dd\dot\phi \, \Gamma(r,\varphi,\theta) g(r,\varphi,\theta,\dot\phi) \, r \cos\varphi \, e^{-ik\theta},\\
\gamma_{1,k,\perp} &\equiv \int_0^\infty \dd r \, r \int_0^{2\pi} \dd\varphi \int_0^{2\pi} \dd\theta \int_{-\infty}^\infty \dd\dot\phi \, \Gamma(r,\varphi,\theta) g(r,\varphi,\theta,\dot\phi) \, r \sin\varphi \, e^{-ik\theta},\\
\gamma_{1,k,\dot\phi} &\equiv \int_0^\infty \dd r \, r \int_0^{2\pi} \dd\varphi \int_0^{2\pi} \dd\theta \int_{-\infty}^\infty \dd\dot\phi \, \Gamma(r,\varphi,\theta) g(r,\varphi,\theta,\dot\phi) \, \dot\phi \, e^{-ik\theta},\\
\tau_{1,k,\parallel} &\equiv \int_0^\infty \dd r \, r \int_0^{2\pi} \dd\varphi \int_0^{2\pi} \dd\theta \int_{-\infty}^\infty \dd\dot\phi \, \tau(r,\varphi,\theta,\dot\phi) g(r,\varphi,\theta,\dot\phi) \, r \cos\varphi \, e^{-ik\theta},\\
\tau_{1,k,\perp} &\equiv \int_0^\infty \dd r \, r \int_0^{2\pi} \dd\varphi \int_0^{2\pi} \dd\theta \int_{-\infty}^\infty \dd\dot\phi \, \tau(r,\varphi,\theta,\dot\phi) g(r,\varphi,\theta,\dot\phi) \, r \sin\varphi \, e^{-ik\theta},\\
\tau_{1,k,\dot\phi} &\equiv \int_0^\infty \dd r \, r \int_0^{2\pi} \dd\varphi \int_0^{2\pi} \dd\theta \int_{-\infty}^\infty \dd\dot\phi \, \tau(r,\varphi,\theta,\dot\phi) g(r,\varphi,\theta,\dot\phi) \, \dot\phi \, e^{-ik\theta}.
\end{align}
\end{subequations}
\end{widetext}
Here, as in the zeroth-order terms, the $\parallel$ and $\perp$ superscripts refer to the components of the collective force $\bm{F}_\text{int}$ parallel and perpendicular to $\hat{\bm{n}}_1$. Respectively, the $\parallel$, $\perp$, and $\dot\phi$ subscripts refer to the components of the spatial gradients parallel and perpendicular to $\hat{\bm{n}}_1$, $\bm{\nabla}_\parallel$ and $\bm{\nabla}_\perp$, and to the spinning-rate gradient $\partial_{\dot\phi}$ in \cref{eq collective-force-torques-order1}.

\subsection{Moment hierarchy and hydrodynamic equations} \label{moment-hierarchy}

To complete the coarse-graining of the microscopic model, we define continuum fields as the angular moments of the one-particle distribution function $\Psi_1$. For example, the zeroth moment corresponds to the density field $\rho(\bm{r},t)$, the first moment corresponds to the polarization density $\bm{p}(\bm{r},t)$, and the second moment is related to the nematic order-parameter tensor density $\bm{Q}(\bm{r},t)$:
\begin{subequations} \label{eq hydrodynamic-fields}
\begin{align}
\rho(\bm{r},t) &=\int \Psi_1(\bm{r},\theta,\dot\phi,t)\, \dd\theta\,\dd\dot\phi,\\
p_\alpha(\bm{r},t) &=\int n_\alpha\, \Psi_1(\bm{r},\theta,\dot\phi,t)\, \dd\theta\,\dd\dot\phi,\\
Q_{\alpha\beta}(\bm{r},t) &=\int \left[n_\alpha n_\beta-\frac{1}{2}\delta_{\alpha\beta}\right] \Psi_1(\bm{r},\theta,\dot\phi,t)\, \dd\theta\,\dd\dot\phi.
\end{align}
\end{subequations}
We can then obtain hydrodynamic equations for each of these fields by taking the corresponding moment of the Smoluchowski equation \cref{eq psi1} for the one-particle distribution $\Psi_1$.
%completed with the collective force and torques given in \cref{eq collective-force-torques-final}. When integrating the terms containing two factors of $\Psi_1$ in \cref{eq collective-force-final}, we assume that they factor out and give rise to nonlinear terms in the hydrodynamic fields. For example, the term that goes like $\Psi_1^2 \hat{\bm n}_1$ in \cref{eq collective-force-final} coarse-grains to a nonlinear term like $\rho {\bm p}$.
We will obtain hydrodynamic equations in this way in the following two sections for the isotropic and the nematic states, respectively.

\section{ISOTROPIC-NEMATIC TRANSITION} \label{isotropic-nematic}

Here, we use our coarse-graining approach to predict the onset of nematic order in our system. In \cref{Fig simulations}, we compare the prediction obtained here with the numerical results from our simulations, described in \cref{simulation-details,simulation-results} above. Given that spin torque transfer has no macroscopic consequences for uniform phases, here we ignore the spin-rate degree of freedom. Thus, our calculation considers self-propelled rolling particles that interact both via central forces and via orientation torques.

Following Refs. \cite{Grossmann2020,Das2024}, to determine a transition between uniform phases, we will consider \cref{eq psi1} to lowest order in gradients:
\begin{equation} \label{eq psi1-zeroth-gradients}
\partial_t \Psi_1 = -\partial_{\theta_1} \frac{\Gamma_\text{int}}{\xi_\text{r}} + D_\text{r}^\text{eff} \partial_{\theta_1}^2 \Psi_1.
\end{equation}
Next, we expand the distribution function in angular Fourier components as
\begin{equation} \label{eq Fourier-expansion-no-spin}
\Psi_1 (\bm{r}',\theta') = \frac{1}{2\pi} \sum_{k=0}^\infty e^{-ik\theta'} \,\tilde{\Psi}_{1,k}(\bm{r}').
\end{equation}
In terms of the Fourier components, \cref{eq psi1-zeroth-gradients} reads
\begin{equation} \label{eq psi1-zeroth-gradients-Fourier}
\partial_t \tilde\Psi_{1,k} = \frac{i k}{2\pi \xi_\text{r}} \sum_{m=0}^\infty \tilde\Psi_{1,m} \gamma_{0,m} \tilde\Psi_{1,k-m} - D_\text{r}^\text{eff} k^2 \tilde\Psi_{1,k},
\end{equation}
where the torque coefficients $\gamma_{0,m}$, first introduced in \cref{eq zeroth-order-coefficients}, are given here by
\begin{equation} \label{eq torque-coefficients}
\gamma_{0,m} = \int_0^\infty \dd r \, r \int_0^{2\pi} \dd\varphi \int_0^{2\pi} \dd\theta \, \Gamma(r,\varphi,\theta) g(r,\varphi,\theta) e^{-im\theta}.
\end{equation}

We next introduce the correspondence of the hydrodynamic fields, defined in \cref{eq hydrodynamic-fields}, with the Fourier components of the one-particle distribution function $\Psi_1$:
\begin{subequations}
\begin{align}
\rho(\bm{r},t) &= \tilde\Psi_{1,0}(\bm{r},t),\\
\bm{p}(\bm{r},t) &= \left( \begin{array}{c} \mathrm{Re} \tilde\Psi_{1,1}(\bm{r},t)\\ \mathrm{Im} \tilde\Psi_{1,1}(\bm{r},t) \end{array} \right),\\
\bm{Q}(\bm{r},t) &= \frac{1}{2} \left( \begin{array}{cc} \mathrm{Re} \tilde\Psi_{1,2}(\bm{r},t) & \mathrm{Im} \tilde\Psi_{1,2}(\bm{r},t) \\
\mathrm{Im} \tilde\Psi_{1,2}(\bm{r},t) & -\mathrm{Re} \tilde\Psi_{1,2}(\bm{r},t) \end{array} \right).
\end{align}
\end{subequations}
Based on these correspondences, we derive the dynamics of the nematic order-parameter tensor $\bm{Q}$ by taking the $k=2$ Fourier mode of \cref{eq psi1-zeroth-gradients-Fourier}, which gives
\begin{multline} \label{eq psi1-2}
\partial_t \tilde \Psi_{1,2} = \frac{2 i}{2\pi \xi_\text{r}} \left(\tilde \Psi_{1,0} \gamma_{0,0} \tilde\Psi_{1,2} + \tilde\Psi_{1,1} \gamma_{0,1} \tilde\Psi_{1,1} \right.\\
\left. + \tilde\Psi_{1,2} \gamma_{0,2} \tilde\Psi_{1,0} \right) - 4D_\text{r} \tilde\Psi_{1,2},
\end{multline}
where we have replaced $D_\text{r}^\text{eff} \rightarrow D_\text{r}$ to account for the fact that reversal events do not impact nematic order, as discussed in \cref{active-chiral}.

We now use the specific orientation torques in the simulations, given in \cref{eq torque-simulations}, as well as the global chiral symmetry in the simulations, which implies $g(r,-\varphi,-\theta) = g(r,\varphi,\theta)$. Introducing these specifications in \cref{eq torque-coefficients}, we obtain the coefficients $\gamma$ in \cref{eq psi1-2}, which are given by
\begin{subequations}
\begin{align}
\gamma_{0,0} &= 0,\\
\gamma_{0,1} & = - i \int_0^{r_\Gamma} \dd r \, r \int_0^{2\pi} \dd\varphi \int_0^{2\pi} \dd\theta \, \Gamma_0 \sin(2\theta) \sin\theta\, g(r,\varphi,\theta),\\
\gamma_{0,2} & = -i \int_0^{r_\Gamma} \dd r \, r \int_0^{2\pi} \dd\varphi \int_0^{2\pi} \dd\theta \, \Gamma_0 \sin^2(2\theta)  g(r,\varphi,\theta). \label{eq gamma_02}
\end{align}
\end{subequations}
With these results, and defining $\gamma_1 \equiv i \gamma_{0,1}$ and $\gamma_2 \equiv i \gamma_{0,2}$ for notation convenience, we recast \cref{eq psi1-2} into the following dynamics for the nematic order-parameter tensor:
\begin{equation}
\partial_t \bm{Q} = \frac{1}{\pi \xi_\text{r}} \left(\gamma_1 \bm{p}\bm{p} + \gamma_2 \rho \bm{Q} \right) - 4D_\text{r} \bm{Q}
\end{equation}
to lowest order in gradients. For states with zero net polarity ($\bm{p}=\bm{0}$), like the ones considered in our simulations, this result further simplifies to
\begin{equation}
\partial_t \bm{Q} =  a_Q [\rho] \, \bm{Q},
\end{equation}
where
\begin{equation} \label{eq nematic-growth-rate-appendix}
a_Q[\rho] = \frac{\gamma_2}{\pi \xi_\text{r}} \rho - 4D_\text{r}
\end{equation}
is the growth rate of the nematic order, as given in \cref{eq nematic-growth-rate}. If $a<0$, the isotropic state with $\bm{Q}=\bm{0}$ is stable. If $a>0$, it is unstable to the appearance of nematic order. Thus, the condition $a_Q[\rho] = 0$ determines the phase boundary for the isotropic-nematic transition.

To predict this phase boundary, we measure the pair distribution function $g(r,\varphi,\theta)$ from our simulations, and we use it to calculate $\gamma_2$ via the following integral:
\begin{equation}
\gamma_2 = \int_0^{r_\Gamma} \dd r \, r \int_0^{2\pi} \dd\varphi \int_0^{2\pi} \dd\theta \, \Gamma_0 \sin^2(2\theta)  g(r,\varphi,\theta).
\end{equation}
Performing this calculation for simulations at different values of the self-propulsion speed $v_0$ and the area fraction $\phi_0$, we evaluate the growth rate in \cref{eq nematic-growth-rate-appendix}, and we look for the values of $v_0$ and $\phi_0$ at which it crosses zero. These crossing values give the phase boundary that we plot in \cref{Fig simulations}. These theoretical results capture the emergence of nematic order observed in our simulations.

\section{HYDRODYNAMIC EQUATIONS FOR THE CHIRAL NEMATIC STATE} \label{nematic-hydrodynamics}

In this section, we use the coarse-graining results of \cref{derivation} to obtain the hydrodynamic equations for the nematic, possibly also chiral, state.

To this end, we first simplify the expressions of the coefficients in \cref{eq zeroth-order-coefficients,eq first-order-coefficients} through three assumptions:
\begin{itemize}
\item{(i)} We introduce the specific forms of the interaction force and torques given in \cref{eq force-torque-fields-expressions}.
\item{(ii)} We assume that the entire system is non-chiral, as half of the screws will spin in one sense and half of the screws in the other. Hence, the pair correlation function has to be invariant upon chiral transformations, i.e. simultaneous change of sign of the angles $\varphi$ and $\theta$, $g(r,-\varphi,-\theta,\dot\phi) = g(r,\varphi,\theta,\dot\phi)$, and upon change of sign of the spinning rate $\dot\phi$, $g(r,\varphi,\theta,-\dot\phi) = g(r,\varphi,\theta,\dot\phi)$.
\item{(iii)} We consider the system of active screws to be deep in the nematic state. In this state, the orientation differences between pairs of particles are very small and, hence, the pair correlation function is sharply peaked around $\theta=0$: $g(r,\varphi,\theta,\dot\phi) \approx g_\text{nem}(r,\varphi,\dot\phi) \delta(\theta)$.
\end{itemize}
With these assumptions, for the zeroth-order coefficients, we get that $\zeta_{0,k}^\parallel$ does not vanish and it is given by
%\begin{subequations}
%\begin{align}
%&\begin{multlined} 
%\zeta_{0,k}^\parallel= \int_0^\infty \dd r \, r \int_0^{2\pi} \dd\varphi \int_{-\infty}^\infty \dd\dot\phi \\
%\times F_r(r) g_\text{nem}(r,\varphi,\dot\phi) \cos\varphi,\end{multlined}\\
%\zeta_0 &= \int_0^\infty \dd r \, r \int_0^{2\pi} \dd\varphi \int_{-\infty}^\infty \dd\dot\phi \, F_r(r) g_\text{nem}(r,\varphi,\dot\phi) \cos\varphi, \label{eq zeta0k}\\
%\zeta_0^\perp &=0,\\
%\gamma_0 &= 0,\\
%\tau_0 &= 0.
%\end{align}
%\end{subequations}
\begin{equation} \label{eq zeta0}
\zeta_0 = \int_0^\infty \dd r \, r \int_0^{2\pi} \dd\varphi \int_{-\infty}^\infty \dd\dot\phi \, F_r(r) g_\text{nem}(r,\varphi,\dot\phi) \cos\varphi,
\end{equation}
Here, we have renamed $\zeta_{0,k}^\parallel \to \zeta_0$ using that this coefficient is now independent of $k$. The other coefficients in \cref{eq zeroth-order-coefficients} vanish: $\zeta_{0,k}^\perp = 0$, $\gamma_{0,k} = 0$, and $\tau_{0,k} = 0$. Therefore, with these specifications for the nematic state, the collective force and torques at zeroth order in the gradient expansion, given in \cref{eq collective-force-torques-order0}, reduce to
\begin{subequations}
\begin{align}
\bm{F}_\text{int}^{(0)} &= -\Psi_1^2 \zeta_0 \hat{\bm{n}}_1,\\
\Gamma_\text{int}^{(0)} &= 0,\\
\tau_\text{int}^{(0)} &= 0.
\end{align}
\end{subequations}

Next, we simplify the expressions of the first-order coefficients (\cref{eq first-order-coefficients}). We obtain
\begin{subequations} \label{eq first-order-coefficients-nematic}
\begin{align}
\zeta_{1,\parallel}^\parallel &= \int_0^\infty \dd r \int_0^{2\pi} \dd\varphi \int_{-\infty}^\infty \dd\dot\phi \, F_r(r) g_\text{nem}(r,\varphi,\dot\phi) \, r^2 \cos^2\varphi,\\
\zeta_{1,\perp}^\perp &= \int_0^\infty \dd r \int_0^{2\pi} \dd\varphi \int_{-\infty}^\infty \dd\dot\phi \, F_r(r) g_\text{nem}(r,\varphi,\dot\phi) \, r^2 \sin^2\varphi,\\
\tau_{1,\dot\phi} &= \int_0^\infty \dd r \int_0^{2\pi} \dd\varphi \int_{-\infty}^\infty \dd\dot\phi \, \tau_r(r) g_\text{nem}(r,\varphi,\dot\phi) \, r \, \dot\phi^2,
\end{align}
\end{subequations}
where we have dropped the subindex $k$ because these coefficients are now independent of the angular mode number. The rest of coefficients in \cref{eq first-order-coefficients} vanish.
%The integrals in the expressions of $\zeta_{1,k,\parallel}^\parallel$ and $\zeta_{1,k,\perp}^\perp$ are different but, for typical forms of the pair correlation function, they take similar values.
For simplicity, following previous work \cite{Zhang2021i}, we replace the integrals in the expressions of $\zeta_{1,k,\parallel}^\parallel$ and $\zeta_{1,k,\perp}^\perp$ by their average, thus replacing both $\zeta_{1,k,\parallel}^\parallel$ and $\zeta_{1,k,\perp}^\perp$ by
\begin{equation} \label{eq zeta1}
\zeta_1 \equiv \frac{1}{2} \int_0^\infty \dd r \int_0^{2\pi} \dd\varphi \int_{-\infty}^\infty \dd\dot\phi \, F_r(r) g_\text{nem}(r,\varphi,\dot\phi) \, r^2.
\end{equation}
Thus, the collective force and torques at first order in the gradient expansion, specified for the nematic state, are given by
\begin{subequations}
\begin{align}
\bm{F}_\text{int}^{(1)} &= -\Psi_1 \zeta_1 \bm{\nabla}\Psi_1,\\
\Gamma_\text{int}^{(1)} &= 0,\\
\tau_\text{int}^{(1)} &= \Psi_1 \tau_{1,\dot\phi} \partial_{\dot\phi} \Psi_1.
\end{align}
\end{subequations}

Putting together the zeroth- and first-order terms, the collective force and torques representing the effects of interactions on a reference particle in the nematic state read
\begin{subequations} \label{eq collective-force-torques-final}
\begin{align}
\bm{F}_\text{int} &= -\Psi_1^2 \zeta_0 \hat{\bm{n}}_1 - \Psi_1 \zeta_1 \bm{\nabla}\Psi_1, \label{eq collective-force-final}\\
\Gamma_\text{int} &= 0,\\
\tau_\text{int} &= \Psi_1 \tau_{1,\dot\phi} \partial_{\dot\phi} \Psi_1.
\end{align}
\end{subequations}

Finally, introducing these results in the Smoluchowski equation (\cref{eq psi1}) and taking the orientational moments as introduced in \cref{moment-hierarchy}, we obtain the following equations for the density, polarity, and nematic order-parameter fields:
\begin{subequations} \label{eq hydrodynamic-equations-before-truncation}
\begin{align} \label{eq density-SI}
&\partial_t \rho = -\bm{\nabla}\cdot\bm{J};\quad \bm{J} = v_\parallel [\rho] \bm{p} - v_\perp \bm{\epsilon}\cdot\bm{p} - \zeta_1 \rho \bm{\nabla}\rho,\\
&\begin{multlined} \label{eq polarity-SI}
\partial_t \bm{p} = - \bm{\nabla}\cdot \left( v_\parallel[\rho] \bm{Q} \right) - \frac{1}{2}\bm{\nabla} \left( v_\parallel[\rho] \rho \right) + \bm{\nabla}\cdot \left( v_\perp \bm{\epsilon}\cdot \bm{Q} \right) \\
+ \frac{1}{2} \bm{\epsilon}\cdot \bm{\nabla} \left( v_\perp \rho \right) + \bm{\nabla}\cdot \left( \zeta_1 \bm{p} \bm{\nabla}\rho \right) - D_\text{r}^\text{eff} \bm{p},
\end{multlined}
%\\
%&\begin{multlined} \label{eq nematic-SI}
%\partial_t Q_{\alpha\beta} = - \partial_\gamma \left( v_\parallel [\rho] T_{\alpha\beta\gamma} \right) 
%+ \delta_{\alpha\beta} \partial_\gamma \left( \frac{v_\parallel[\rho]}{2} p_\gamma \right) \\
%+ \partial_\gamma \left( v_\perp \epsilon_{\gamma\mu} T_{\alpha\beta\mu} \right)
%-\partial_\gamma \left( \frac{v_\perp}{2} \delta_{\alpha\beta} \epsilon_{\gamma\mu} p_\mu \right)\\
%+ \partial_\gamma \left( \zeta_1 Q_{\alpha\beta} \partial_\gamma \rho \right) - D_\text{r} Q_{\alpha\beta},
%\end{multlined}
\end{align}
\end{subequations}
which are quoted in \cref{eq density,eq polarity}. To obtain these equations, when integrating the terms containing two factors of $\Psi_1$ in \cref{eq collective-force-final}, we assume that they factor out and give rise to nonlinear terms in the hydrodynamic fields. For example, the term that goes like $\Psi_1^2 \hat{\bm n}_1$ in \cref{eq collective-force-final} coarse-grains to a nonlinear term like $\rho {\bm p}$. The hydrodynamic equations \cref{eq hydrodynamic-equations-before-truncation}, together with the definitions
\begin{subequations} \label{eq hydrodynamic-parameter-definitions}
\begin{align}
v_\parallel [\rho] &= \omega \ell_\parallel - \zeta_0 \rho,\\
v_\perp &= \omega \ell_\perp,\\
D_\text{r}^\text{eff} &= D_\text{r} + 2f_\text{rev},
\end{align}
\end{subequations}
and the results for $\zeta_0$ and $\zeta_1$ in \cref{eq zeta0,eq zeta1}, constitute the final result of our derivation for the nematic state. They specify the collective behavior of the system of active screws in terms of the parameters of their microscopic model, complemented with the pair correlation function in the nematic state, $g_\text{nem}$, needed to obtain $\zeta_0$ and $\zeta_1$.
%\ra{Note that the whereas the decay rate of the polarity in \cref{eq polarity-SI} is the effective rotational diffusivity $D_\text{r}^\text{eff} = D_\text{r} + 2f_\text{rev}$, the decay rate of the nematic order tensor in \cref{eq nematic-SI} is the bare rotational diffusivity $D_\text{r}$. This difference reflects that directional reversals change the polarity but not the nematic order.}

In general, the hydrodynamic equation for the $k^{\text{th}}$ moment involves the $k+1^{\text{th}}$ moment, giving rise to a hierarchy of hydrodynamic equations. In \cref{eq hydrodynamic-equations-before-truncation}, the polarity equation contains the nematic order parameter tensor $\bm{Q}$, which is the second-order orientational moment of $\Psi_1$ as defined in \cref{eq hydrodynamic-fields}.
%the equation for the nematic order parameter contains the third-rank orientational tensor
%\begin{equation}
%T_{\alpha\beta\gamma}(\bm{r},t)=\int n_\alpha n_\beta n_\gamma \, \Psi_1(\bm{r},\theta,\dot\phi,t)\,\dd\theta\,\dd\dot\phi,
%\end{equation}
%which is the third-order orientational moment of $\Psi_1$. Here, we consider only the first three equations of the hierarchy (\cref{eq hydrodynamic-equations-before-truncation}).

\section{FLOW FIELDS AROUND TOPOLOGICAL DEFECTS} \label{fit}

We analyze the flows of \textit{M. xanthus} bacteria around topological defects reported in Ref.~\cite{Han2025} (\cref{Fig plus-experiments,Fig minus-experiments}). Using the procedure in Ref.~\cite{Han2025}, we fit the flows predicted by an incompressible active nematic, but now extended to include the chiral active stress $\zeta_\text{c} \bm{\epsilon}\cdot \bm{Q}$ (see \cref{eq velocity}). Setting other parameter values as in Ref.~\cite{Han2025}, we fit the achiral and chiral active stress coefficients, $\zeta$ and $\zeta_\text{c}$, respectively, divided by the isotropic friction coefficient $\xi_0$. We obtain $\zeta/\xi_0 = 0.46\pm 0.01$ $\mu$m$^2$/min and $\zeta_\text{c} = 0.05 \pm 0.01$ $\mu$m$^2$/min for $+1/2$ defects, and $\zeta = 0.72 \pm 0.01$ $\mu$m$^2$/min and $\zeta_\text{c} = -0.17 \pm 0.01$ $\mu$m$^2$/min for $-1/2$ defects. These values are used to generate the theory plots in \cref{Fig defects}.

\end{document}